%% file: main.tex
\documentclass[fleqn,usenatbib]{mnras}

\usepackage{newtxtext,newtxmath}
\usepackage{soul}
\usepackage{textgreek}
\usepackage[T1]{fontenc}

\DeclareRobustCommand{\VAN}[3]{#2}
\let\VANthebibliography\thebibliography
\def\thebibliography{\DeclareRobustCommand{\VAN}[3]{##3}\VANthebibliography}

\usepackage{graphicx}	% Including figure files
\usepackage{amsmath}	% Advanced maths commands

\def\M87{M87$^{\ast}$}
\def\m87{M87$^{\ast}$}
\def\sgra{Sgr\,A$^\ast$}

\def\MIseventeen{EHTC\,\M872017\,{I}}
\def\MIIseventeen{EHTC\,\M872017\,{II}}
\def\MIIIseventeen{EHTC\,\M872017\,{III}}
\def\MIVseventeen{EHTC\,\M872017\,{IV}}
\def\MVseventeen{EHTC\,\M872017\,{V}}
\def\MVIseventeen{EHTC\,\M872017\,{VI}}
\def\MVIIseventeen{EHTC\,\M872017\,{VII}}
\def\MVIIIseventeen{EHTC\,\M872017\,{VIII}}
\def\MIXseventeen{EHTC\,\M872017\,{IX}}
\def\MMWseventeen{EHTC\,MWL\,\M872017}
\def\MIeighteen{EHTC\,\M872017\,{I}}
\def\MIIeighteen{EHTC\,\M872017\,{II}}

\def\SIseventeen{EHTC\,\sgra2017\,{I}}
\def\SIIseventeen{EHTC\,\sgra2017\,{II}}
\def\SIIIseventeen{EHTC\,\sgra2017\,{III}}
\def\SIVseventeen{EHTC\,\sgra2017\,{IV}}
\def\SVseventeen{EHTC\,\sgra2017\,{V}}
\def\SVIseventeen{EHTC\,\sgra2017\,{VI}}

\defcitealias{EHTC2019I}{{\MIseventeen}}
\defcitealias{EHTC2019II}{{\MIIseventeen}}
\defcitealias{EHTC2019III}{{\MIIIseventeen}}
\defcitealias{EHTC2019IV}{{\MIVseventeen}}
\defcitealias{EHTC2019V}{{\MVseventeen}}
\defcitealias{EHTC2019VI}{{\MVIseventeen}}
\defcitealias{EHTC2021VII}{{\MVIIseventeen}}
\defcitealias{EHTC2021VIII}{{\MVIIIseventeen}}
\defcitealias{EHTC2023IX}{{\MIXseventeen}}
\defcitealias{EHTMWL2021}{{\MMWseventeen}}
\defcitealias{EHTC2023I}{{\MIeighteen}}
\defcitealias{EHTC2023II}{{\MIIeighteen}}
\defcitealias{EHTC2022I}{{\SIseventeen}}
\defcitealias{EHTC2022II}{{\SIIseventeen}}
\defcitealias{EHTC2022III}{{\SIIIseventeen}}
\defcitealias{EHTC2022IV}{{\SIVseventeen}}
\defcitealias{EHTC2022V}{{\SVseventeen}}
\defcitealias{EHTC2022VI}{{\SVIseventeen}}

\def\YuanNarayan{Yuan\,\&\,Narayan\,2014}
\defcitealias{YuanNarayan2014ARA&A..52..529Y}{{\YuanNarayan}}

\def\KATHENA{K-ATHENA}
\defcitealias{Grete2021ITPDS..32...85G}{{\KATHENA}}
\def\Athena{Athena++}
\defcitealias{WhiteAthena++2023ApJ...949..103W}{{\Athena}}
\def\BHAC{BHAC}
\defcitealias{Porth_2017_BHAC}{{\BHAC}}
\def\COSMOS{COSMOS++}
\defcitealias{Anninos_2005_COSMOS++}{{\COSMOS}}
\def\ECHO{ECHO}
\defcitealias{Del_Zanna_2007_ECHO}{{\ECHO}}
\def\HAMR{H-AMR}
\defcitealias{Liska2022ApJS..263...26L}{{\HAMR}}
\def\KHARMA{KHARMA}
\defcitealias{Prather2021JOSS....6.3336P}{{\KHARMA}}
\def\IllinoisGRMHD{IllinoisGRMHD}
\defcitealias{Etienne_2015_IllinoisGRMHD}{{\IllinoisGRMHD}}
\def\KORAL{KORAL}
\defcitealias{Sadowski2014_KORAL}{{\KORAL}}
\def\Uwabami{Uwabami}
\defcitealias{Takahashi_2016ApJ_Uwabami}{{\Uwabami}}

\input{Title/Title}

\begin{document}

\label{firstpage}
\pagerange{\pageref{firstpage}--\pageref{lastpage}}
\maketitle

%%%%%%%%%%%%%%%%%%%% ABSTRACT %%%%%%%%%%%%%%%%%%%%

\input{0Abstract/Abstract}

%%%%%%%%%%%%%%%%%%%%%%%%%%%%%%%%%%%%%%%%%%%%%%%%%%

%%%%%%%%%%%%%%%%% BODY OF PAPER %%%%%%%%%%%%%%%%%%

\input{1Intro/Intro}

\input{2Simulations/Simulations}

\input{3Disk/Disk}

\input{4Jet/Jet}

\input{6Conclusions/Conclusions}
\section*{Acknowledgements}

We gratefully recognise Sasha Tchekhovskoy and Doosoo Yoon for insightful discussions. %as well as the Black Hole Initiative at Harvard University, via its Distinguished Visitors Program. %and the Gravitation and Astroparticle Physics Amsterdam (GRAPPA) Institute at the University of Amsterdam. 
L.S and S.M. were supported by a Dutch Research Council (NWO) VICI award, grant No. 639.043.513 and by a European Research Council (ERC) Synergy Grant "BlackHolistic" grant No. 101071643. In addition, L.S was supported by the Colfuturo Scholarship. G.M. was supported by a Canadian Institute of Theoretical Astrophysics (CITA) postdoctoral fellowship and by a Netherlands Research School for Astronomy (NOVA), Virtual Institute of Accretion (VIA) postdoctoral fellowship. K.C. was supported in part by grants from the Gordon and Betty Moore Foundation and the John Templeton Foundation to the Black Hole Initiative at Harvard University, and by NSF award OISE-1743747. O.P. acknowledges funding from VIA within NOVA. ML was supported by the John Harvard, ITC and NASA Hubble Fellowship Program fellowships. B.R. is supported by the Natural Sciences \& Engineering Research Council of Canada (NSERC) and by a grant from the Simons Foundation (MP-SCMPS-00001470). Research at the Flatiron Institute is supported by the Simons Foundation.

This research was enabled by support provided by grant no. NSF PHY-1125915 along with a INCITE program award PHY129, using resources from the Oak Ridge Leadership Computing Facility, Summit, which is a US Department of Energy office of Science User Facility supported under contract DE-AC05- 00OR22725, as well as Calcul Quebec (http://www.calculquebec.ca) and Compute Canada (http://www.computecanada.ca). The computational resources and services used in this work were partially provided by facilities supported by the Scientific Computing Core at the Flatiron Institute, a division of the Simons Foundation; and by the VSC (Flemish Supercomputer Center), funded by the Research Foundation Flanders (FWO) and the Flemish Government – department EWI. This research is part of the Frontera \citep{Stanzione2020Frontera:Foundation} computing project at the Texas Advanced Computing Center (LRAC-AST20008). Frontera is made possible by National Science Foundation award OAC-1818253. This work used the Dutch national e-infrastructure with the support of the
SURF Cooperative using grant no. EINF-3036 and EINF-5383, which is (partly) financed by the Dutch Research Council (NWO), for post-processing of simulation data.
%\textcolor{magenta}{enter your spider grant number here} 
%The Acknowledgements section is not numbered. Here you can thank helpful colleagues, acknowledge funding agencies, telescopes and facilities used etc. Try to keep it short.

%%%%%%%%%%%%%%%%%%%%%%%%%%%%%%%%%%%%%%%%%%%%%%%%%%
\section*{Data Availability}

The simulation post-processed data used to plot the images in this work are available in Zenodo at http://doi.org/10.5281/zenodo.10996780

%Example from Koushik's paper Data used to plot the images in this work are available in Zenodo at http://doi.org/10.5281/zenodo.5044837.



\bibliographystyle{mnras}
\bibliography{main} % if your bibtex file is called example.bib

%\begin{thebibliography}{99}
%\end{thebibliography}

%%%%%%%%%%%%%%%%%%%%%%%%%%%%%%%%%%%%%%%%%%%%%%%%%%

%%%%%%%%%%%%%%%%% APPENDICES %%%%%%%%%%%%%%%%%%%%%

%\input{7Appendix/Appendix}

%%%%%%%%%%%%%%%%%%%%%%%%%%%%%%%%%%%%%%%%%%%%%%%%%%

% Don't change these lines
\bsp	% typesetting comment
\label{lastpage}
\end{document}

%% file: Title/Title.tex
\title[Resolution MAD]{Resolution analysis of magnetically arrested disk simulations}

\author[L. Salas et al.]{
L. Salas$^{1}$\thanks{E-mail: l.d.sosapantasalas@uva.nl}
G. Musoke$^{2,1}$
K. Chatterjee$^{3,4}$
S.B. Markoff$^{1,5}$
O. Porth$^{1}$
M.T.P. Liska$^{6,7}$
B. Ripperda$^{2,8-11}$
\\
$^{1}$Anton Pannekoek Institute for Astronomy, University of Amsterdam, Science Park 904, 1098 XH Amsterdam, The Netherlands\\
$^{2}$Canadian Institute for Theoretical Astrophysics, University of Toronto, 60 St. George Street, Toronto, ON M5S 3H8, Canada\\
$^{3}$Institute for Research in Electronics and Applied Physics, University of Maryland, 8279 Paint Branch Drive,
College Park, MD 20742, USA\\
$^{4}$Black Hole Initiative at Harvard University, 20 Garden Street, Cambridge, MA 02138, USA\\
$^{5}$Gravitation and Astroparticle Physics Amsterdam Institute, University of Amsterdam, Science Park 904, 1098 XH 195 196 Amsterdam, The Netherlands\\
$^{6}$Center for Relativistic Astrophysics, Georgia Institute of Technology, Howey Physics Bldg, 837 State St NW, Atlanta, GA, 30332, USA\\
$^{7}$Institute for Theory and Computation, Harvard University, 60 Garden Street, Cambridge, MA 02138, USA\\
$^{8}$Dunlap Institute for Astronomy and Astrophysics, University of Toronto, 50 St. George Street, Toronto, ON M5S 3H4, Canada\\
$^{9}$Department of Physics, University of Toronto, 60 St. George Street, Toronto, ON M5S 1A7, Canada\\
$^{10}$Perimeter Institute for Theoretical Physics, 31 Caroline Street North, Waterloo, ON N2L 2Y5, Canada\\
$^{11}$Center for Computational Astrophysics, Flatiron Institute, 162 5th avenue, New York, NY, 10010, USA\\
}

\date{Accepted 2024 July 25. Received 2024 July 19; in original form 2024 April 19}

\pubyear{2024}

%% file: 0Abstract/Abstract.tex
\begin{abstract}
Polarisation measurements by the Event Horizon Telescope from \M87 and \sgra  suggest that there is a dynamically strong, ordered magnetic field, typical of what is expected of a magnetically arrested accretion disk (MAD). In such disks the strong poloidal magnetic field can suppress the accretion flow and cause episodic flux eruptions. Recent work shows that General Relativistic Magnetohydrodynamic (GRMHD) MAD simulations feature dynamics of turbulence and mixing instabilities that are becoming resolved at higher resolutions. We perform a convergence study of MAD states exceeding the status quo by an order of magnitude in resolution. We use existing 3D simulations performed with the H-AMR code, up to resolution of 5376 × 2304 × 2304 in a logarithmic spherical-polar grid. We find consistent time-averaged disk properties across all resolutions. However, higher resolutions reveal signs of inward angular momentum transport attributed to turbulent convection, particularly evident when mixing instabilities occur at the surfaces of flux tubes during flux eruptions. Additionally, we see wave-like features in the jet sheath, which become more prominent at higher resolutions, that may induce mixing between jet and disk. At higher resolutions, we observe the sheath to be thinner, resulting in increased temperature, reduced magnetisation, and greater variability. Those differences could affect the dissipation of energy, that would eventually result in distinct observable radiative emission from high-resolution simulations. With higher resolutions, we can delve into crucial questions about horizon-scale physics and its impact on the dynamics and emission properties of larger-scale jets.

\end{abstract}

\begin{keywords}
accretion, accretion disks -- black hole physics -- MHD
\end{keywords}

%Accretion is a fundamental astrophysical process, occurring across all scales of black hole mass. Despite its ubiquitous nature, the accretion process, alongside its connection to jet outflows, poses many fundamental questions. General Relativistic Magneto hydrodynamic (GRMHD) simulations are providing significant insights into the nature of black hole accretion and jet outflows. Following recent efforts in the Event Horizon Telescope (EHT) collaboration to compare numerical solutions between different GRMHD codes, we now aim to perform a convergence study between five simulations conducted using the GPU-accelerated GRMHD code H-AMR, up to a resolution of $5375 \times 2304 \times 2304$ in a logarithmic spherical-polar grid. The objective is to assess how well global properties in the disk and jet agree for different resolutions, whether there is any essential physics missing at low resolution and what the appropriate resolution threshold is for obtaining consistent results.
%The goal is to determine the level of agreement between simulations, examine the alteration in the disk and jet evolution, and determine the consistency of the overall results across all resolutions considered. 
%\textcolor{red}{\textit{In the end, sketch the results instead of the aim}}
%It should be a single paragraph, not more than 250 words (200 words for Letters).

%%%%%%%%%%%%%%%%%%%%%%%%%%%%%%%%%%%%%%%%%%%%%%%%%%

%% file: 1Intro/Intro.tex
\section{Introduction}

The accretion of matter is a fundamental astrophysical process occurring across a variety of compact objects such as neutron stars, white dwarfs, and black holes (BHs). Jets and wind outflows are a natural consequence of the accretion process and are present across all scales of black hole mass-- from black hole X-ray binaries to the supermassive black holes powering the bright compact regions at the centres of most massive galaxies (active galactic nuclei; AGN). Accretion disks and outflows emit radiation across the electromagnetic spectrum, from radio frequencies to high energy gamma rays \citepalias[e.g.][]{EHTMWL2021,EHTC2022II}. The relativistic outflows (jets) launched by AGN in particular are thought to have a profound effect on their environments, playing a key role in regulating star formation, galaxy evolution and the evolution of galaxy clusters \citep[e.g.][]{silk1998quasars,schawinski2007observational,fabian2012observational}. Numerical simulations have improved our understanding of the accretion process and the mechanisms by which jets are launched and affect their environments. Yet it remains unclear which resolutions are sufficient to accurately capture the global evolution of accreting black holes, and jets.
%The mechanisms of how jets affect their environments are still under active investigation. How exactly jets affect their environments and the mechanisms by which they do so are still under active investigation. Improving our understanding of the accretion process is a fundamental step towards addressing the many open questions concerning black holes and the influence they have on their surroundings.

Theoretical modelling and numerical simulations, combined with multi-wavelength observations have provided critical insights into the nature of outflows (jets) and radiation from accreting black hole systems \citepalias[e.g.][]{YuanNarayan2014ARA&A..52..529Y,EHTC2019I,EHTC2022I}. In particular, GRMHD simulations have become increasingly successful in explaining the dynamics of the accretion flow, winds, and the evolution of relativistic jets close to the black hole, therefore playing a fundamental role in the interpretation of observational data. %For instance, simulations conducted by \cite{Balbus_Hawley_1998} demonstrated that magnetohydrodynamical (MHD) turbulence, the so called Magneto-Rotational Instability (MRI), can drive accretion by angular momentum transport in magnetized fluids. %Later in GRMHD simulations, \cite{hawley2004general} showed a generic structure of the accretion flow with the disk, the corona, the funnel, and the funnel wall jet. 
%General Relativistic Magneto-hydrodynamic (GRMHD) simulations are helpful to investigate the impact of different physical configurations (e.g. magnetic field strength, black hole spin, thermal properties) into the structure of magnetized accretion flows and jets, therefore playing a fundamental role in the interpretation of observational data.
%\citep{hawley2004general,hawley2006magnetically,McKinney2012MNRAS.423.3083M} 
GRMHD simulations have been significantly improving over the last few decades owing to the increase in computing power and the development of efficient numerical algorithms in multiple GRMHD codes \citepalias[e.g.][]{WhiteAthena++2023ApJ...949..103W, Grete2021ITPDS..32...85G,Porth_2017_BHAC,Anninos_2005_COSMOS++,Del_Zanna_2007_ECHO,Liska2022ApJS..263...26L,Prather2021JOSS....6.3336P,Etienne_2015_IllinoisGRMHD,Sadowski2014_KORAL,Takahashi_2016ApJ_Uwabami}.
%(e.g. K-ATHENA \citealt{Grete2021ITPDS..32...85G}, BHAC \citealt{Porth_2017_BHAC}, COSMOS++ \citealt{Anninos_2005_COSMOS++}, ECHO \citealt{Del_Zanna_2007_ECHO}, 
%HARM \citealt{Gammie2003_HARM}, H-AMR \citealt{Liska2022ApJS..263...26L}, 
%iharm3D, KHARMA (Prather2021JOSS....6.3336P) IllinoisGRMHD \citealt{Etienne_2015_IllinoisGRMHD}, KORAL \citealt{Sadowski2014_KORAL}, Uwabami \citealt{Takahashi_2016ApJ_Uwabami}). 
\cite{EHTGRMHDcomparison} %used a set of diagnostic quantities to 
demonstrated the general robustness of GRMHD simulation results regardless of the different algorithms, implementations, and grid geometries of different numerical codes in the weakly magnetized regime (SANE).
Recently, there has been significantly more interest in a category of numerical solutions wherein strong poloidal magnetic fields can interrupt the accretion flow, forming a magnetically arrested disk \citep[MAD;][]{Bisnovatyi-Kogan1974Ap&SS..28...45B,Narayan2003PASJ...55L..69N}.
\citet{MADresWhite} conducted a resolution study of MAD accretion flows, with a maximum effective 173 cells per decade in radius, 256 cells in polar angle and 512 cells in azimuthal angle. They showed that the general large-scale structure of the flow is robust with resolution. However, the spatial structure and Lorentz factor of the jet were not fully converged. 

%Ideal GRMHD simulations utilise several underlying assumptions: (1) The accretion disk is treated as a fully ionised plasma with infinite electrical conductivity, thus resistivity and electric fields are neglected. (2) The mean free path for Coulomb scattering of electrons and ions is large compared to the size of the black hole's event horizon. (3) The exchange of energy between the plasma and the radiation field is ignored. (4) The system of governing equations is closed by an ideal gas equation of state, i.e. non-ideal effects such as viscosity are neglected \citep[see e.g.][]{Gammie2003_HARM,Ripperda2019ApJS..244...10R}. Despite the ideal approximation, 
%(A background spacetime must be specified , for the disk-jet systems of interest in this work, the fixed spacetime is the rotating Kerr black hole metric. 
%On the other hand, the strong poloidal magnetic fields present in magnetically arrested disks 

In recent years, the unprecedented resolution of the Event Horizon Telescope (EHT) has enabled the first direct imaging of the plasma surrounding the supermassive black holes \M87 and \sgra \citepalias{EHTC2019I,EHTC2022I}. %of \M87 and \sgra
The EHT observations revealed a bright ring, formed by the hot luminous plasma close to the black hole, along with a dark central region, the black hole 'shadow'. 
%These 'event horizon-scale' observations, combined with theoretical modelling and numerical simulations, contributed significant insight into the nature of the accretion process around black holes \citepalias[e.g.][]{EHTC2019V, EHTC2022V}. 
%and provided critical tests of the theory of General relativity. 
%The EHT observations provided constraints on the size of the black hole shadow of \M87 and \sgra and the shape of the ring, %the inclination of the source, variability characteristics and the bolometric luminosity, and have also corroborated the mass measurement obtained from alternative techniques \citep[e.g.][]{Gebhardt2011ApJ...729..119G,GRAVITYCollaboration2020A&A...636L...5G}. 
Synthetic images produced by ray-tracing of GRMHD simulations are the fiducial models used by the EHT collaboration to compare against observations. %The shadow size and the appearance of the ring were found to be consistent with the those predicted by General Relativity \citepalias[e.g.][]{EHTC2019VI, EHTC2022VI}. 
The closest-matching GRMHD simulations to the EHT observations suggest that both \M87 and \sgra exhibit accretion flows in the MAD regime \citepalias[e.g.][]{EHTC2019V, EHTC2021VIII, EHTC2023IX, EHTC2022V}. Moreover, \sgra can reach the MAD state in the inner region when poloidal magnetic fields are advected inwards from larger scales, as demonstrated by wind-fed accretion simulations \citep{Ressler2023MNRAS.521.4277R}. Additionally, \citet{Liska2020MNRAS.494.3656L} showcased that turbulence within the accretion disk itself can generate strong poloidal magnetic fields in-situ.

Recent work shows that MAD states feature complicated dynamics that become resolved at higher resolutions. Consequently, simulating MAD states is numerically challenging. Episodic flux eruptions result in the ejection of large cavities containing flux tubes with low-density through magnetic reconnection \citep{Ripperda2020ApJ...900..100R,Ripperda2022BlackReconnection}, which may serve as a mechanism for flare generation \citep{DexterTchekhovskoy2020MNRAS.497.4999D,Porth2021MNRAS.502.2023P,Ripperda2022BlackReconnection,Hakobyan2023ApJ...943L..29H}. The Rayleigh-Taylor instability (RTI) mechanism acting on the boundary of a flux tube can lead to secondary instabilities, turbulence, and magnetic reconnection \citep{Zhdankin2023PhRvR...5d3023Z}. \cite{Ripperda2022BlackReconnection} captured plasmoid-mediated magnetic reconnection in a 3D MAD state for the first time, by conducting a GRMHD simulation with maximum effective resolution of 1630 cells per decade in radius, 2304 cells in polar angle and 2304 cells in azimuthal angle. Plasmoid-mediated magnetic reconnection can result in particle energisation through dissipation and may explain flares from accreting black holes, in particular the TeV flares observed from \M87 \citep{Aharonian2006Sci...314.1424A,Acciari2010ApJ...716..819A,Aliu2012ApJ...746..141A,blanch2021magic}, and the infrared and X-ray flares thought to originate close to the event horizon of \sgra \citep{Baganoff2001Natur.413...45B,Eckart2004A&A...427....1E,neilsen2015x,GRAVITY2018A&A...618L..10G,abuter2021constraining}.

The current simulations conducted with the Graphics Processing Unit (GPU)-accelerated GRMHD code H-AMR \citep{Liska2022ApJS..263...26L} are pushing the boundaries of what is computationally feasible with regards to resolution. However, conducting high resolution GRMHD simulations is computationally extremely expensive. %H-AMR's speed performance has enabled high resolution and long-duration simulations. %thus it is not feasible to conduct all numerical studies at these extreme resolutions. %Yet it remains unclear which resolutions are sufficient to accurately capture the global evolution of accreting black holes. 
In this paper, we perform a convergence study on a wider range of resolutions than has ever been previously explored (see Table~\ref{tab:Sims}), employing existing 3D ideal GRMHD simulations with H-AMR of an accreting black hole in the MAD regime \citep{Ripperda2022BlackReconnection}. We determine how well the physical properties of the accretion flow at different spatial resolutions agree. %We explore the effects of spatial resolution on the physical properties of the disk and jets, %determine how the evolution of mass accretion rate and magnetic flux are influenced by increasing the accuracy in which the MRI is resolved, 
We identify the processes that are only recovered at high resolutions. We constrain the minimum resolution %, of those that we test, 
that ensures consistent and converged results, which
%Determining the resolutions at which the simulations begin to converge 
will enable the GRMHD community to minimise the computational cost of these simulations, without sacrificing the accuracy of the global evolution of the disk-jet-black hole system.

The structure of this paper is as follows: Section~\ref{sec:Sims} provides the numerical setup for simulations and an overview of the H-AMR code. The impact of resolution on the disk and the jet is analysed in Section~\ref{sec:Results}. The conclusions of the study are presented in Section~\ref{sec:Conclusions}. %and ~\ref{sec:Conclusions}.

%% file: 2Simulations/Simulations.tex
\section{Simulations}
\label{sec:Sims}

\subsection{GRMHD equations}
\label{sec:GRMHD}

The equations of GRMHD comprise the particle number conservation equation:
\begin{equation}
     \left ( nu^{\mu} \right )_{;\mu} = 0\text{ ,}
     \label{eq:number_conservation_equation}
\end{equation}
the energy momentum conservation equations:
\begin{equation}
     {T^{\mu}}_{\nu;\mu} = 0\text{ ,}
     \label{eq:energy_momentum_conservation_equation}
\end{equation}
and the Maxwell's equations:
\begin{equation}
     {F^{\ast\mu\nu}}_{;\nu} = 0\text{ ,}
     \label{eq:Maxwell_equation}
\end{equation}
In ideal MHD, the dual of the electromagnetic ﬁeld
tensor is $F^{\ast\mu\nu}=b^{\mu}u^{\nu}-b^{\nu}u^{\mu}$ and the stress-energy tensor is:
\begin{equation}
    {T^{\mu}}_{\nu}\equiv \left (\rho+u_g+p_g+b^2 \right ) u^{\mu} u_{\nu}-b^{\mu} b_{\nu}+\left (\frac{1}{2}b^2+p_g \right ){\delta^{\mu}}_{\nu} \text{ ,}
	\label{eq:Tud}
\end{equation}
Here $n$ is the particle number density, $\rho=mn$ is the rest mass density\footnote{Normalised to a maximum density $\rho _{\textup{max}}=1$.}, $m$ is the mean rest mass per particle, $u_g$ is the gas energy density, $p_g$ is the gas pressure, and $b^2/2$ is the magnetic energy density. The 4-velocity and the magnetic field 4-vector are $u^{\mu}$ and $b^{\mu}$ respectively. ${\delta^{\mu}}_{\nu}$ is the Kronecker delta.
%In ideal GRMHD simulations, the dynamics of the plasma around the BH are sensitive to 
The gas pressure is proportional to the fluid-frame rest-mass density and proton temperature ($p_g\propto \rho T_p $) and the magnetic pressure $p_b=b^2/2$, where $b$ is the magnetic field strength in the frame co-moving with the fluid. The proton temperature ($T_p$), in CGS units:
\begin{equation}
T_p = \frac{ m_p c^2 p_g}{k_B\rho} \text{ ,}
\label{eq:Tp}
\end{equation}
%where $\mu = 1.69$ is the mean molecular weight %\op{more like the ion temperature then?  I think we always put mu=1, so this is a bit confusing. Please rewrite this first paragraph, it puts the wrong emphasis imo. LS: you're right}
where $m_p$ is the proton mass and $k_B$ is the Boltzmann constant. We use geometrized units with gravitational constant, black hole mass, and speed of light $G=M=c=1$, and a factor of $1/\sqrt{4\pi}$ is absorbed in the deﬁnition of the magnetic ﬁeld. Greek indices run through $[0,1,2,3]$ and Roman indices through $[1,2,3]$. The metric determinant is $g$.

\subsection{H-AMR implementation}
\label{sec:H-AMR}

H-AMR \citep{Liska2022ApJS..263...26L}, a state-of-the-art 3D GRMHD code, builds upon the foundation of the original HARM2D code \citep{Gammie2003_HARM}. H-AMR converts conserved quantities, such as particle number density and energy-momentum density, into primitive variables, including rest-mass density, internal energy density, and velocity components, utilising the inversion scheme by \citet{Noble2006}. The Adaptive mesh refinement (AMR), local adaptive time-stepping (LAT), and GPU acceleration within a hybrid CUDA-OpenMP-MPI framework utilised in H-AMR speed-up computations. The achieved resolutions are $\approx10\times10\times5$ times higher than those in the previous MAD study \citep{MADresWhite}. %by 2-5 orders of magnitude for a wide range of applications, enabling particularly challenging problems to be simulated at unprecedented resolutions. For example, the LAT $\times$ GPU $\times$ SMR (Static Mesh Reﬁnement) speed-up factor is approximately $200$ for resolution R5 (Table~\ref{tab:Sims}) as compared against a 20-core Skylake CPU without LAT and with uniform grid. For challenging problems such as tilted thin accretion disks and simulating jets at large distances from the black hole, the AMR speed-up factor is $\sim $10–1000 \citep{Liska2022ApJS..263...26L}.
%5 orders of magnitude when you multiply all speed factors: LAT × GPU × SMR  × AMR for the ﬁducial run of a tilted thin accretion disk in Liska+2022

%\op{without LAT and with uniform grid.}
%\op{200 compared to what? LS: problem solved}.

H-AMR utilises a staggered grid for constrained transport of magnetic fields as described in \citep{Gardiner2005JCoPh}, and solves the GRMHD equations of motion in conservative form in arbitrary (fixed) spacetimes. H-AMR uses a finite volume, shock-capturing Godunov-based HLLE scheme, with third order accurate spatial reconstruction of cell variables (PPM, \citealp{Colella1984}) on cell faces and second order accurate time evolution. The simulations are performed on a logarithmic spherical-polar grid in a Kerr-Schild foliation, $\log(r)$, $\theta$ and $\phi$. Since cells get squeezed near the pole, the timestep in all spherical grids is reduced by an additional factor proportional to the resolution in the $\phi$-direction. To remedy this issue, %the cells are stretched out immediately adjacent to the pole in the $\theta$-direction \citep{Tchekhovskoy2011EfficientHole} and 
multiple levels of static mesh de-refinement in the $\phi$-direction are used to keep the aspect ratio of the cells close to uniform at high latitudes. This method prevents the squeezing of cells near the pole from reducing the global timestep, while maintaining high accuracy in all three dimensions (see section 3.4 in \citealp{Liska2022ApJS..263...26L}). The simulations presented in this work utilise outflow boundary conditions in the radial direction, transmissive boundary conditions in the $\theta$-direction and periodic boundary conditions in the $\phi$-direction.

\subsection{Numerical setup}
\label{sec:setup}

In order to study convergence in GRMHD simulations of MAD accretion flows around a spinning black hole, we analyse a total of five 3D GRMHD simulations. The simulation with the highest resolution, R5, is performed with an effective\footnote{To prevent the squeezing of cells, three internal and four external derefinement levels in $\phi$ are used to reduce the resolution from $N_{\phi}=128-2304$ at $30^{\circ}<\theta < 150^{\circ}$ to $N_{\phi}=16-18$ within $0.5^{\circ}-7.5^{\circ}$ of each pole \citep{Liska2022ApJS..263...26L}.} resolution of $5376 \times 2304 \times 2304$ and the other four simulations we present are a factor $\approx 2 \times 2 \times 2$ less in resolution (see Table~\ref{tab:Sims}). The radial domain of the simulation grid is $r=\left [ 1.2 - 2000 \right ]r_g$. Each simulation is evolved for at least $10^4r_g/c$ (see fifth column of Table~\ref{tab:Sims}). In all runs the disk is initialised using a torus in hydrostatic equilibrium \citep{Fishbone1976RelativisticHoles} around a Kerr black hole with dimensionless spin $a=0.9375$. The inner edge of the torus is located at $r=20 r_g$ and the pressure maximum at $r=41 r_g$. The torus is threaded with a single poloidal magnetic field loop, defined by the $\phi$-component of the vector potential $A_{\phi} \propto  \textup{max}\left [ \rho/\rho _{\textup{max}}\left ( r/r_\textup{in} \right )^3 \sin ^3\theta \exp\left ( -r/400 \right )-0.2 , 0\right ]$, and normalised to obtain a gas-to-magnetic-pressure ratio $p_{g,\mathrm{max}}/p_{b,\mathrm{max}}=100$. We adopt an equation of state for an ideal gas with an adiabatic index of $\gamma_{\mathrm{ad}} = 13/9$, which corresponds approximately to a gas mixture of relativistic electrons $\gamma_{\mathrm{ad},e} = 4/3$ and non-relativistic ions $\gamma_{\mathrm{ad},p} = 5/3$. A semi-relativistic gas with $\gamma_{\mathrm{ad}}\sim 1.55$ was found for BH accreting at $\sim 10^{-6}$ the Eddington accretion rate, in radiative two-temperature GRMHD simulations \citep{Liska2024ApJ...966...47L}. Within the spine region the following floor and ceiling values are employed; the rest-mass density floor is $\rho _{\textup{fl}}=\mathrm{MAX}[b^2/25,10^{-7}r^{-2},10^{-20}]$, the gas energy density floor is $u _{g,\textup{fl}}=\mathrm{MAX}[b^2/750,10^{-9}r^{-26/9},10^{-20}]$ and the magnetisation ceiling is $\sigma _{\textup{max}}=25$ where $\sigma=b^2/\rho$. The disk is initialised with the same energy density perturbation for all resolutions, $u_g(1 + 0.04(\textrm{rand([0, 1])} - 0.5))$, where $\textrm{rand([0, 1])}$ is a random deviate between 0 and 1. %\citep[e.g.][]{fragile2009general}.

\begin{table}
	\centering
	\caption{Number of cells in the radial $N_{r}$, polar $N_{\theta}$ and azimuth $N_{\phi}$ directions for each simulation, together with the final time of the simulation $t_{\mathrm{sim}}$.} %\textcolor{red}{GM: Please replace the 'size' column with Lundquist number. Also add a Column for max evolution time for each run}}
	\label{tab:Sims}
	\begin{tabular}{ccccc} % four columns, alignment for each
		\hline
		Resolution & $N_{r}$ & $N_{\theta}$ & $N_{\phi}$ & $t_{\mathrm{sim}} (r_g/c)$\\
		\hline
		R1 & 288 & 128 & 128 & 20420\\
		R2 & 580 & 288 & 256 & 22790\\
		R3 & 1280 & 576 & 512 & 12330\\
		R4 & 2240 & 1056 & 1024 & 10780\\
		R5 & 5376 & 2304 & 2304 & 10000\\
		\hline
	\end{tabular}
\end{table}

%\textcolor{red}{GM: Add here other GRMHD resolution studies numerical algorithm studies like Siegel + 2018 https://iopscience.iop.org/article/10.3847/1538-4357/aabcc5/pdf and refs therein. LS: Done}

%In the set of GRMHD equations, conserved variables (e.g. particle number density and energy-momentum density) and primitive variables (e.g rest-mass density, internal energy density, velocity components) are defined. Initially, \cite{Gammie2003_HARM} proposed a Newton-Raphson scheme to convert conserved to primitive variables in a set of five non-linear equations. Later, \citet{Noble2006} reduced the problem to two non-linear equations improving the numerical performance. Current GRMHD codes are based on the recommended two dimensional Noble scheme. For the ideal gas equation of state, \citet{Siegel2018ApJ...859...71S} found no significant improvement in speed and accuracy among the different schemes.

%H-AMR \citep{Liska2022ApJS..263...26L} is a massively parallel 3D GRMHD code based originally on the GRMHD HARM2D code \citep{Gammie2003_HARM}. H-AMR converts conserved (e.g. particle number density and energy-momentum density) to primitive variables (e.g rest-mass density, internal energy density, velocity components) using the inversion scheme of \citet{Noble2006}. H-AMR has been extensively modified from these original sources in order to increase the code's speed through numerous features including adaptive mesh refinement (AMR), local adaptive time-stepping (LAT) and GPU acceleration in a natively developed hybrid CUDA-OpenMP-MPI framework.

%% file: 3Disk/Disk.tex
\section{Results}
\label{sec:Results}

There are various definitions of jets and unbound outflows in the literature \citep[see e.g.][]{Narayan2012GRMHDOutflows,yuan2015numerical}. One option to define the jet region is by using $\left (-{T^{r}}_{t} /(\rho u^r) \right )^2-1>1$, where $-{T^{r}}_{t}$ is the radial ($\mu=1$) - temporal ($\nu=0$) component of the stress-energy tensor (Equation~\ref{eq:Tud}), representing the outward radial energy flux. This is equivalent to the region where the total energy per unit rest-mass is greater than $\sqrt{5}c^2$ \citepalias{EHTC2019V,EHTC2021VIII}. Alternatively the jet region can also be defined as in \citet{moscibrodzka2013coupled}, where the jet is taken to be the unbound gas outﬂowing with a minimum bulk velocity $v_{min}/c = 0.2$ as measured in a normal observer frame, and the jet spine is taken to be the region at which $\sigma > 0.1$. In this paper, we follow the definitions in \citet{davelaar2018general} in Kerr-Schild coordinates, where the jet is characterised by two components: (1) The jet spine-- a relativistic and strongly magnetized outflow ($\sigma > 1$). (2) The jet sheath-- a mildly relativistic outflow defined using the Bernoulli parameter $Be = -\bar{h} u_t>1.02$ where $\bar{h}=(\rho+u_g+p_g)/\rho$ is the specific gas enthalpy and $u_t$ is the time component of the inverse four-velocity.
%\op{I think this comes from Monika originally, check and cite appropriately please.}

\begin{figure*}
	\includegraphics[width=\textwidth]{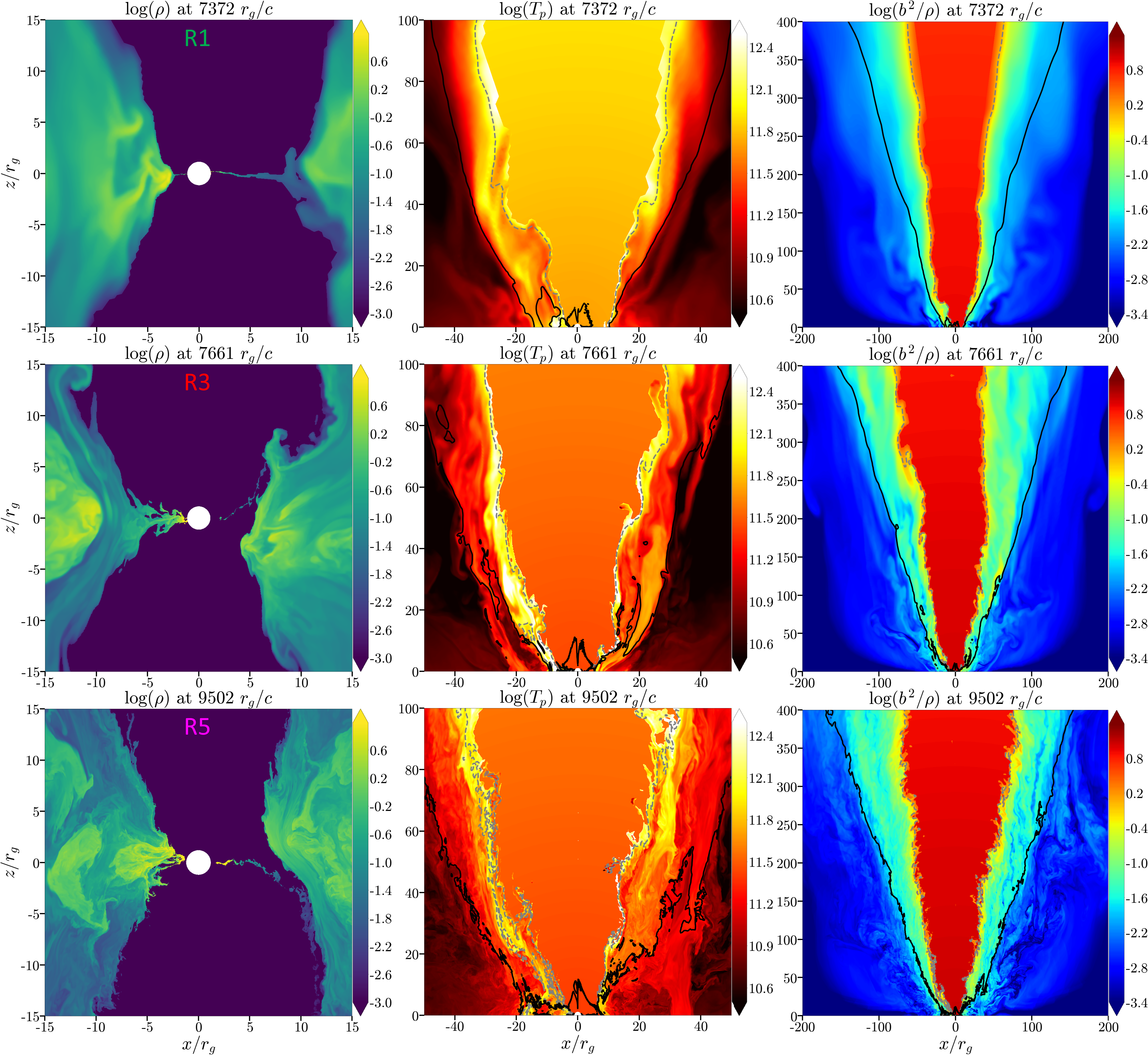}
    \caption{Meridional 2D slices through $\phi=0$ for (log) density $\rho$ (left column), proton temperature $T_p$ in Kelvin (middle column), and magnetisation $\sigma=b^2/\rho$ (right column) during a flux eruption event, when the normalised magnetic flux on the event horizon $\varphi _ {BH}$ approaches a local minimum (Fig.~\ref{fig:fluxesZoom}). From top to bottom: resolution R1, R3 and R5. The grey dashed and black lines correspond to surfaces where $\sigma = 1$ and $Be=1.02$, respectively. The jet has two components: the spine (defined as the region in which $\sigma > 1$) and the sheath (defined as the region in which $\sigma < 1$ and $Be>1.02$). The dominance of numerical floors undermines the reliability of the jet spine. We partially cover the spine with a zero screen for the plot of $\rho$ and time-average screens for $T_p$ and $\sigma$. These screens are activated for $\sigma > 3$, allowing visibility of plasma within the range of $1 < \sigma < 3$.}
    \label{fig:rho512}
\end{figure*}

For a qualitative perspective, Fig.~\ref{fig:rho512} shows 2D slices through azimuthal angle $\phi=0$ for density $\rho$, proton temperature $T_p$, and magnetisation $\sigma = b^2/\rho$ for the simulations conducted at different resolutions. %The electron temperature is calculated in post-processing using Equation~\ref{eq:T_p/T_e} \citep{moscibrodzka2016general}, see Section~\ref{sec:TpTe}. 
%\op{What prescription do you use to get the electron temperature?}
As the simulation resolution increases, wave-like features in the jet sheath become more prominent and plasmoids in current sheets close to the event horizon are captured for resolution R5. Additionally, the jet sheath gets hotter and less magnetised.
%A color map of the density is shown in Fig.~\ref{fig:rho512} for $t=10^4 r_g/c$. In the zoom in panels of the left is visible the artificial injection of matter in the polar regions. As the resolution is increased, more finer details are captured.

\subsection{Disk properties}
\label{sec:Disk}

%We use the following notation to indicate integrating over $\theta$ and $\phi$
%\begin{equation}
%    \left \{ X \right \}_{\rho[\theta,\phi]}^{\textrm{disk}} = \iint X \rho \left ( \rho > 10^{-4} \right ) \left ( \sigma < 1 \right ) \left ( Be < 1.02 \right )\sqrt{-g}d\theta d\phi \text{ .}
%	\label{eq:Average}
%\end{equation}

In order to determine quantitative differences between the simulations we compare various fluid parameters, as defined in the following subsections. The averaged profile of a variable $X$ is calculated by integrating over $\theta$ and $\phi$ with a combination of three conditions to select the disk; the disk is taken to be the region that satisfies density $\rho/\rho _{\textup{max}} = \rho > 10^{-4}$, magnetisation $\sigma < 1$ and Bernoulli parameter $Be<1.02$. Additionally, we include a density weight to give more relevance to regions of the disk with higher density. \footnote{Note the difference in notation, $\left \langle  \right \rangle$ without any sub/super script is time average.}
\begin{equation}
    \left \langle X \right \rangle_{\rho[\theta,\phi]}^{\textrm{disk}} = \frac{\iint X \rho \left ( \rho > 10^{-4} \right ) \left ( \sigma < 1 \right ) \left ( Be < 1.02 \right )\sqrt{-g}d\theta d\phi}{\iint \rho \left ( \rho > 10^{-4} \right ) \left ( \sigma < 1 \right ) \left ( Be < 1.02 \right )\sqrt{-g}d\theta d\phi} \text{ .}
	\label{eq:AverageDisk}
\end{equation}

\subsubsection{Fluxes}
\label{sec:AccretionRates} % used for referring to this section from elsewhere
In this section we determine the impact of resolution on the temporal evolution of the mass, magnetic and energy fluxes in the simulations. The fluxes are defined as follows\footnote{Subscripts indicate where in the radial domain the fluxes are analysed, e.g. $\dot{M}_{BH}$ at the event horizon and $\dot{M}_{5r_g}$ at $5r_g$.}.
%the double integral in the polar and azimuth directions ($\theta$ and $\phi$ respectively) of the appropriate variable. Thus the fluxes across concentric shells depend only on radius. 
The mass accretion rate $\dot{M}$ is given by:
\begin{equation}
    \dot{M}\equiv -\iint \rho u^r\sqrt{-g}d\theta d\phi \text{ .}
	\label{eq:Mdot}
\end{equation}
$\dot{M} _{BH}$ is contaminated by artificial matter injection due to density floors above and below the black hole poles up to a radius of $\approx 5r_g$. Therefore, $\dot{M} _{5r_g}$ is a better representation of the real accretion process. The energy flux $\dot{E}$ is given by:
\begin{equation}
    \dot{E}\equiv -\iint {T^{r}}_{t}\sqrt{-g}d\theta d\phi \text{ .}
	\label{eq:Edot}
\end{equation}
%The angular momentum flux $\dot{L}$ is given by: \textcolor{red}{GM: why was this removed? LS: Because I didn't use $\dot{L}$ in the analysis, should I add that plot?}
%\begin{equation}
%    \dot{L}\equiv -\iint {T^{r}}_{\phi}\sqrt{-g}d\theta d\phi \text{ .}
%	\label{eq:Ldot}
%\end{equation}
The energy outflow efficiency of the jet spine is the energy return rate divided by the time averaged mass accretion rate:
\begin{equation}
   \eta_j\equiv \frac{\dot{M}-\dot{E}_{\mathrm{spine}}}{\left \langle \dot{M} \right \rangle} \text{ .}
	\label{eq:eta}
\end{equation}
To calculate the energy flux in the jet spine only $\dot{E}_{\mathrm{spine}}$, we add an extra condition before doing the integral, which is the magnetisation greater than 1 ($\sigma > 1$). The magnetic flux is defined as:
\begin{equation}
   \Phi\equiv \frac{1}{2}\iint \left | B^r \right |\sqrt{-g}d\theta d\phi
	\label{eq:Phi} \text{ .}
\end{equation}
To study MAD simulations is convenient to use the normalised magnetic flux:
\begin{equation}
   \varphi\equiv \frac{\Phi}{\sqrt{ \left \langle \dot{M} \right \rangle}} \text{ ,}
	\label{eq:phi}
\end{equation}
known as the “MAD parameter” which, for spin $a=0.9375$ and torus scale height $H/R \approx 0.3$, has the critical value $\varphi _{max}\approx 15$ (within the units adopted here; \citet{tchekhovskoy2012general}). $H$ and $R$ are the full height and cylindrical radius of the disk, respectively.

\begin{figure}
	\includegraphics[width=\columnwidth]{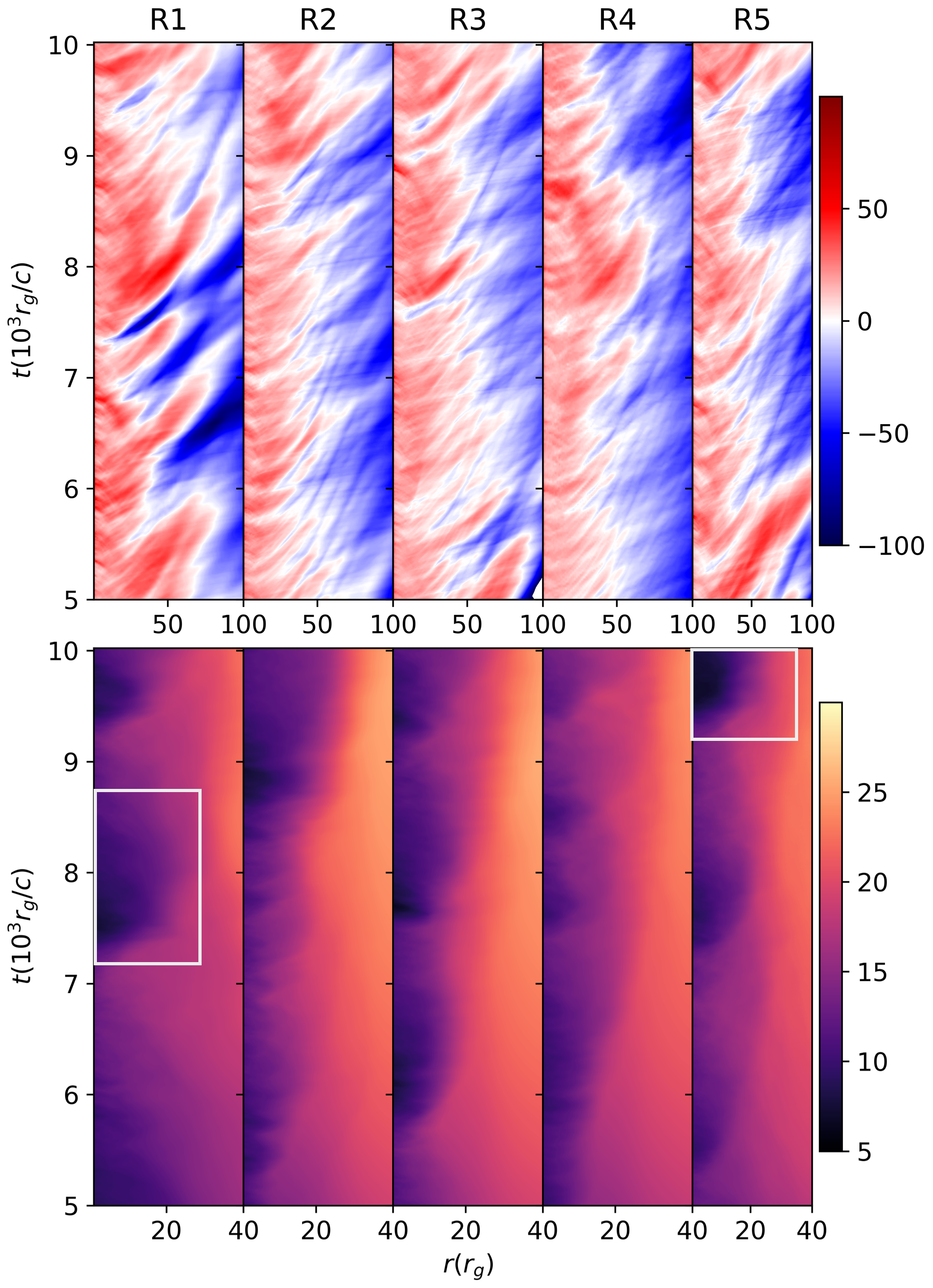}
    \caption{Space-time diagrams of the mass accretion rate (top panel) and normalised magnetic flux (bottom panel). From left to right: resolution R1 to R5. In MAD states, the magnetic flux on the event horizon can build up until it is saturated and gets ejected. In these episodic flux eruption events, the magnetic pressure interrupts the accretion flow, causing flux tubes of low-density material with vertical field to be pushed into the disk until a radius in between $15$ and $40 r_g$ %\op{20 or 30?} 
    (e.g. see the white boxes on the bottom panels). Two major eruption events occur for R1, several small eruptions for R2 and R4 and three eruptions for R3 and R5. This behaviour is seen as well in the plot of normalised magnetic flux versus time in the middle panels of Fig.~\ref{fig:fluxesZoom}.}
    %\op{IMO this is a very important matter which should be discussed carefully.  Also I'd like to know the answer, not sure if you can be conclusive in this study.  To me it looks more like you have few eruptions at low res, then more at mid res and again few at high res.  There is not a lot of statistics to go on as you only have a handful events in the 5k M.  }
    %\textcolor{red}{GM: state here what changes you actually see with resolution. For the Mdot plots maybe reduce the outer radius so that it is consistent with the region you show in the magnetic flux panel. }}
    \label{fig:MdotMagFlux}
\end{figure}

Space-time diagrams of the mass accretion rate (Equation~\ref{eq:Mdot}) and normalised magnetic flux (Equation~\ref{eq:phi}) are presented in Fig. ~\ref{fig:MdotMagFlux} for the five resolutions of Table~\ref{tab:Sims}. A quasi-stationary state of the mass flux is obtained after $t=5\times 10^3 r_g/c$ until a radius of around $50 r_g$. For $r\approx 50r_g$, there is a balance between inflow and outflow so $\left \langle \dot{M} _{50r_g} \right \rangle_{[5-10]\times 10^3 r_g/c} \approx 0$, that marks the point where the torus begins to spread outwards due to viscosity. The relativistic jet and outflows launched from the inner disk have short dynamical timescales. At $\approx80 r_g$, the negative $\dot{M}$ is predominantly due to unbound outflows (spine and sheath). At $\approx100$ and $200 r_g$, unbound outflows contribute $50\%$ and $10\%$ of the total $\dot{M}$, respectively. For example, $\left \langle \dot{M} _{100r_g} \right \rangle_{[5-10]\times 10^3 r_g/c} \approx -40$ in code units, with equal contribution from unbound outflows and viscous spreading of the disk. 
%\op{$100 r_g$ strikes me as very large for a short runtime of just 10k M.  Also your Figures 5+9 indicates there is a stationary state \textbf{in the disk} maybe out to 30 rg.  You might want to indicate the Mdot=0 curve on the Figure where the torus begins to spread outwards due to viscosity.  I'm more optimistic about the jet since that has a very short dynamical time and the environment around the jet is fed by material launched from the inner disk which has short viscous timescale.  So one can justify the 100 for the jet more easily.}
In the analysis of averaged profiles, the properties of the simulations are calculated as time averaged between $t=\left [ 8 - 10 \right ]\times 10^3 r_g/c$ to avoid the dependency on the initial conditions for $r>100 r_g$ and $t<8\times 10^3 r_g/c$. The disk properties still depend on the initial conditions of the Fishbone-Moncrief torus for radii larger than $40 r_g$.
%\textcolor{red}{GM:the space-time diagrams are not time averaged, time is variable. LS: I hope it's clear what I'm trying to say now. That's why I prefer to keep the Mdot plots until an outer radius of $200 r_g$}

%Lorentz factor:

%\begin{equation}
%   \alpha\equiv \frac{1}{\sqrt{-g^{t t}}}u^t
%	\label{eq:alpha}
%\end{equation}

% \begin{figure}
% 	% To include a figure from a file named example.*
% 	% Allowable file formats are eps or ps if compiling using latex
% 	% or pdf, png, jpg if compiling using pdflatex
% 	\includegraphics[width=\columnwidth]{Figures/phibh50_5-10.png}
%     \caption{Spacetime diagrams of the magnetic flux as a function of time and radius. From left to right: resolution R1 to R5, respectively. In MAD, when a saturation level is reached in the event horizon of the BH, the magnetic field is so strong that it can interrupt the accretion flow, causing discrete blobs of material to penetrate the disrupted region and involve sporadic flux eruption events until a radius of $\approx 30 r_g$.}
%     \label{fig:MagFlux}
% \end{figure}

\begin{figure}
	\includegraphics[width=\columnwidth]{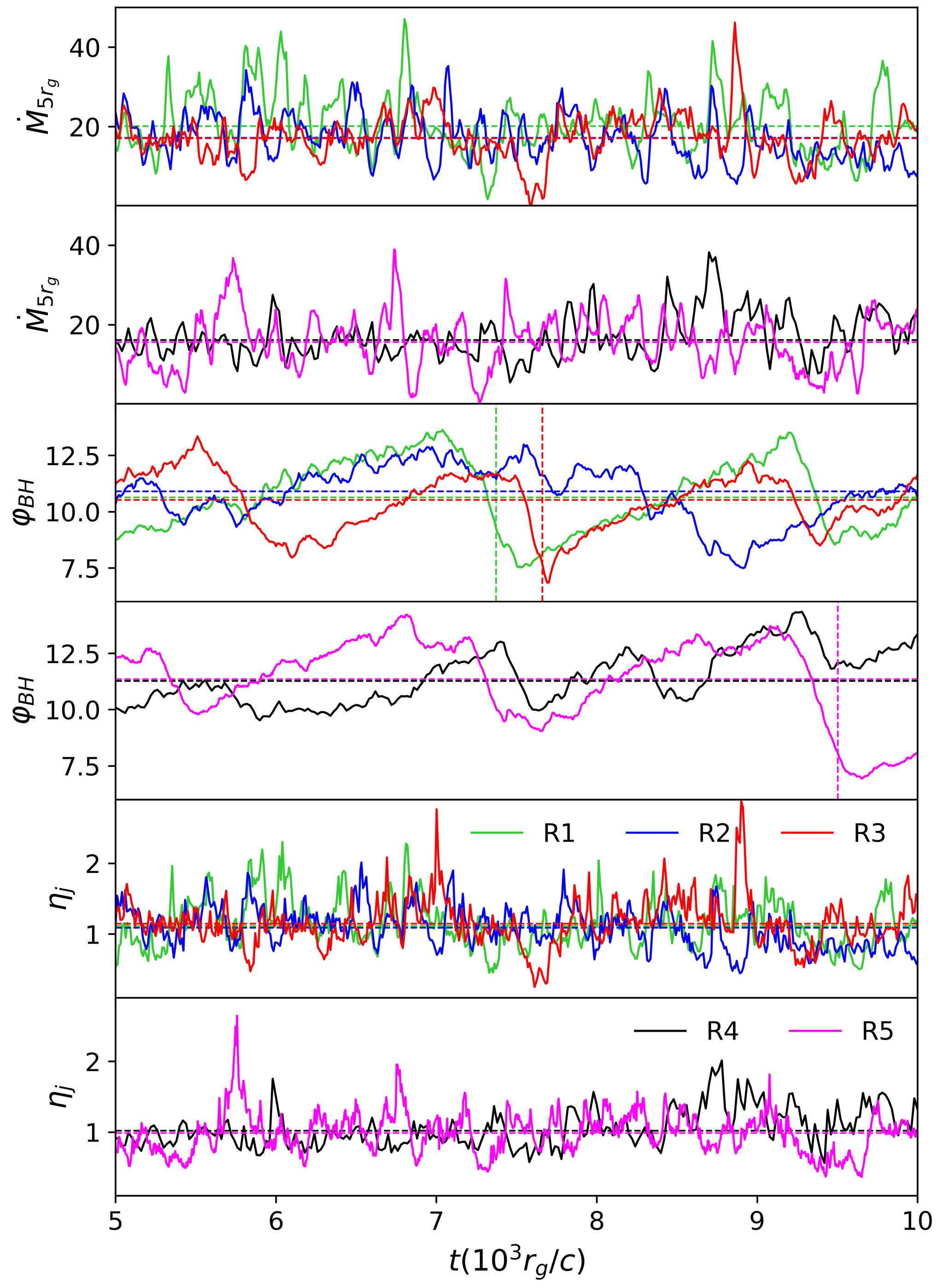}
    \caption{Mass accretion rate, normalised magnetic flux and jet outflow efficiency as a function of time in solid lines, together with the time averaged values in horizontal dashed lines. The magnetic flux saturates after $t=5\times 10^3 r_g/c$. The vertical dashed lines indicate the points where  $\varphi _ {BH}$ approaches a local minimum (Fig.~\ref{fig:rho512}). The colour convention in all figures is resolution R1: green, R2: blue, R3: red, R4: black, and R5: purple.}
    \label{fig:fluxesZoom}
\end{figure}

\begin{figure}
	\includegraphics[width=\columnwidth]{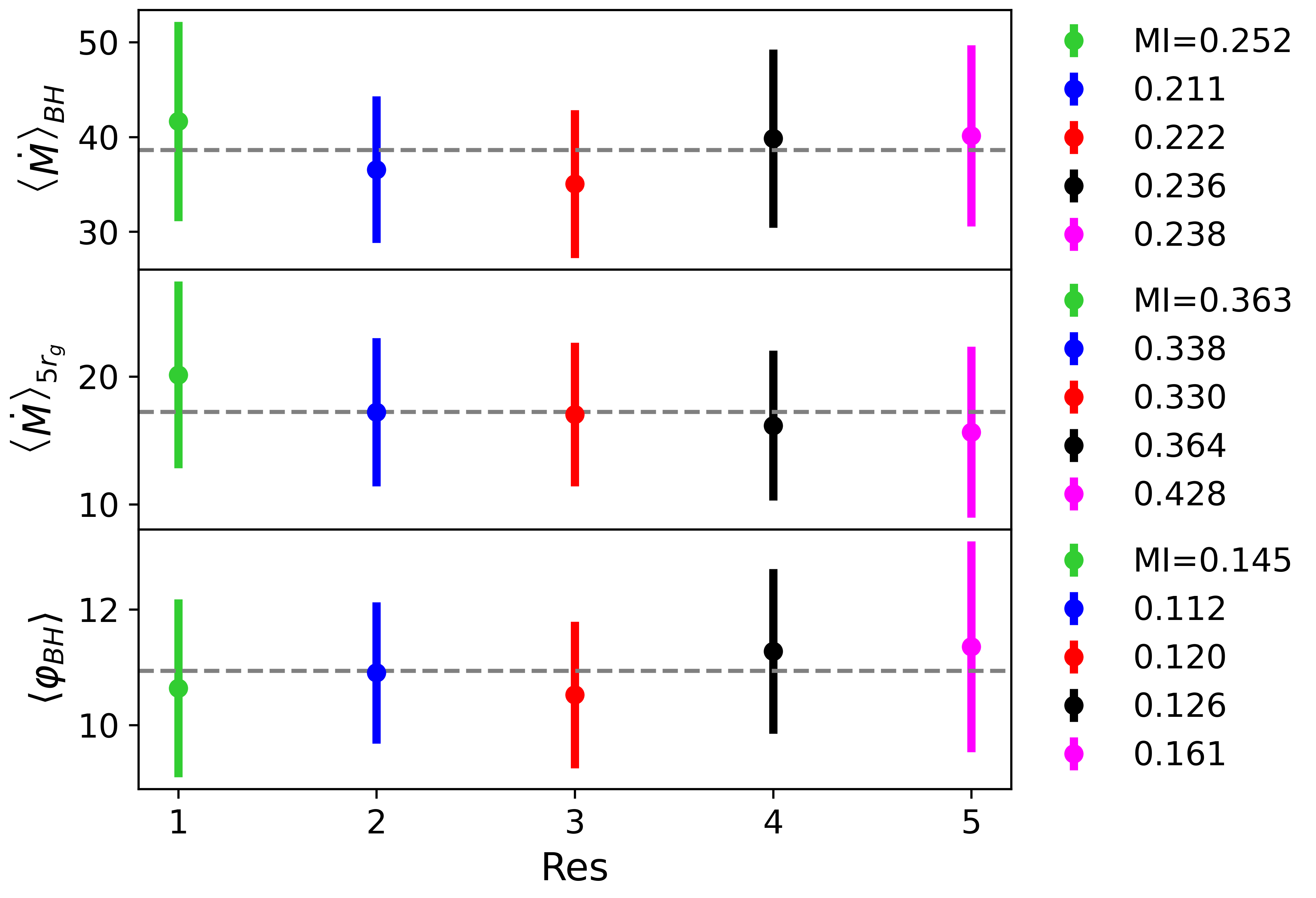}
    \caption{Resolution dependence of average values and standard deviations of mass accretion rate at the event horizon (top panel) and at $r=5r_g$ (middle panel) and normalised magnetic flux (bottom panel). The horizontal grey line in each panel is the average value of all five resolutions, respectively. The modulation index is shown on the right. %\op{what errorbars are these? The standard error of the mean? Please indicate.}
    The error bars represent the standard deviations. Calculation interval $t=[5-10]\times 10^{3} r_g/c$.}
    \label{fig:fluxes_var}
\end{figure}

The time evolution of the mass accretion rate $\dot{M}$ at $5r_g$, the normalised magnetic flux $\varphi$ and the jet outflow
efficiency $\eta_j$ 
%\textcolor{red}{GM: write out what these are in words too} 
at the event horizon are presented in Fig.~\ref{fig:fluxesZoom}. The magnetic flux grows until saturation level at around $t=5\times 10^3 r_g/c$. The fluctuations are caused by accumulation and escape of field line bundles in the vicinity of the BH. The magnetic pressure of those escaping field lines stops the accretion process for a short time, therefore, a drop in the magnetic flux is correlated to a sudden decrease of the accretion rate. %The average value of $\eta_j$ is greater than 1 and can reach values of almost 3, indicating that there is a net energy flow tapped from a spinning BH via the Blandford–Znajek mechanism \citep{Tchekhovskoy2011EfficientHole}. Note the large fluctuations in energy outflow efficiency $\eta_j$, are not exactly correlated with corresponding fluctuations in $\phi_{BH}$.
We find that the mass accretion rate decreases with increasing resolution when $\dot{M}$ is computed at a radius of $5r_g$. However, we note that there is no clear tendency for $\dot{M}$ when computed at the horizon-- this is due to the artificial matter injection that occurs close to the poles of the BH due to flooring (see Fig.~\ref{fig:fluxes_var} and Table.~\ref{tab:variability}).

The peak, average value and standard deviation of $\dot{M}$%at the event horizon and at $5r_g$
, $\varphi$ and $\eta_j$ are presented in Table~\ref{tab:variability}. We use the modulation index $\textup{MI}$ as a standardised measure of time variability of the accretion rates. It is defined as the ratio of the standard deviation to the mean. In Fig.~\ref{fig:fluxes_var}, the $\textup{MI}$ of $\dot{M}_{5r_g}$ initially decreases from resolution R1 to R3, and from R3 it increases. Similarly, $\textup{MI}$ of $\varphi _ {BH}$ increases from R2. 
%\op{notation: there is the MAD parameter $\phi$ and the horizon penetrating flux $\Phi_{\rm BH}$, please be consistent, I don't know what you mean by $\phi_{\rm BH}$.  LS: I added the definition of $\phi_{\rm BH}$ in the previous paragraph}
%This behavior can be explained as the MRI is better resolved with higher resolution, the fluctuations in the fluid are more significant and the magnetic flux eruptions are more frequent. 
However, the differences in $\textup{MI}$ are not statistically significant. The accuracy of the $\textup{MI}$ calculation is dependent on the length of the simulation time window of analysis, $t=[5-10]\times 10^{3} r_g/c$. Longer simulations can provide a more precise measure of the variability. 
Additionally at higher resolutions, we find greater variability of proton temperature, magnetisation and opening angle of the jet sheath close to the black hole at $\approx40r_g$ (see Section~\ref{sec:Jet}). All these differences could affect the predicted  multiwavelength spectra, relevant to the EHTC analysis of \sgra where the two most promising MAD models show higher light-curve variability that historical observations at 230 GHz \citepalias{EHTC2022V}.

\begin{table*}
	\centering
	\caption{Peak, average value and standard deviation of the mass accretion rate at the event horizon and at $5r_g$, the normalised magnetic flux, and the energy outflow efficiency of the jet spine. Calculation interval $t=[5-10]\times 10^{3} r_g/c$.}
	\label{tab:variability}
	\begin{tabular}{ccccccccc} % four columns, alignment for each
		\hline
		Res & $\dot{M}_{BH,max}$ & $\left \langle \dot{M}_{BH} \right \rangle$ & $\dot{M}_{5r_g,max}$ & $\left \langle \dot{M}_{5rg} \right \rangle$ & $\varphi_{BH,max}$ & $\left \langle \varphi_{BH} \right \rangle$ & $\eta_{j,max}$ & $\left \langle \eta_{j} \right \rangle$\\
		\hline
		R1 & 75.9 & 41.6 $\pm$ 10.5 & 47.1 & 20.1 $\pm$ 7.3 & 13.6 & 10.6 $\pm$ 1.5 & 2.30 & 1.12 $\pm$ 0.33 \\
		R2 & 59.2 & 36.6 $\pm$ 7.7  & 35.2 & 17.2 $\pm$ 5.8 & 13.0 & 10.9 $\pm$ 1.2 & 2.01 & 1.09 $\pm$ 0.29 \\
		R3 & 70.0 & 35.0 $\pm$ 7.8  & 46.2 & 17.0 $\pm$ 5.6 & 13.3 & 10.5 $\pm$ 1.3 & 2.87 & 1.15 $\pm$ 0.33 \\
		R4 & 65.0 & 39.8 $\pm$ 9.4  & 38.2 & 16.2 $\pm$ 5.9 & 14.3 & 11.3 $\pm$ 1.4 & 2.01 & 1.02 $\pm$ 0.28 \\
		R5 & 72.7 & 40.1 $\pm$ 9.5  & 38.9 & 15.7 $\pm$ 6.7 & 14.2 & 11.4 $\pm$ 1.8 & 2.64 & 0.99 $\pm$ 0.29 \\
		\hline
	\end{tabular}
\end{table*}

\citet{Tchekhovskoy2011EfficientHole} ran simulations with resolution $288 \times 128 \times \left \{ 32,64,128 \right \}$, seeded with large poloidal magnetic ﬂux to achieve the MAD state, enabling the production of efficient outflows. The authors demonstrated the extraction of net energy from a spinning black hole's ergosphere, via the Penrose–Blandford–Znajek mechanism \citep{blandford1977electromagnetic}.
The authors found an increase of the average energy outﬂow efﬁciency from $\left \langle \eta_j \right \rangle\approx 0.30$ to $\approx 1.40 \pm 0.15$ when the BH spin is increased from $a=0.5$ to $0.99$, respectively. \citet{McKinney2012MNRAS.423.3083M} extended those results by conducting a parametric study of MAD flows with a similar resolution of $272 \times 128 \times 256$ around a rotating black hole with different values of spin, poloidal and toroidal magnetic field geometries. \citet{McKinney2012MNRAS.423.3083M} confirmed that the Poynting flux outflow is powered by the Blandford-Znajek mechanism when the poloidal magnetic flux saturates near the BH, independently of the initial
poloidal seed. Additionally the authors found outflow efficiencies $\left \langle \eta_j \right \rangle\gtrsim 1$ if $\left | a \right |\gtrsim 0.9$. The simulations we present in this work are conducted with $a=0.9375$, and we find outflow efficiencies $\left \langle \eta_j \right \rangle=\left [0.99-1.15\right ]\pm 0.30$. We find that the outflow efficiency is converged within statistical uncertainty for the five resolutions considered (Table~\ref{tab:variability}), where resolution R1 is similar to those used in \citet{Tchekhovskoy2011EfficientHole,McKinney2012MNRAS.423.3083M}. Additionally, we find that $\eta_j$ fluctuates with time (Fig. ~\ref{fig:MdotMagFlux}) and has maximum values up to $2.87$ (Table~\ref{tab:variability}).
%evidencing a net energy extraction from a spinning BH \citep{blandford1977electromagnetic}.

%For resolution R1, there are regions in the disk close to the BH and around $100 r_g$ where the MRI-driven turbulence is not well sustained (see Section~\ref{sec:MRI}).
%\op{This text seems out of place here and disconnected from the stuff above. LS: You're right, removed}

\subsubsection{Flux eruptions}
\label{sec:FluxEruptions}

The variation in the amount of major flux eruptions observed across different resolutions, as depicted in Fig.~\ref{fig:MdotMagFlux} and \ref{fig:fluxesZoom}, raises two key points: 1) Periodicity may not be directly linked to resolution, and 2) variations in periodicity suggest that it might not be inherently indicative of a periodic physical phenomenon. However, the statistics of these findings are limited by the relatively short duration of our high resolution simulations \citep[see e.g.][]{Narayan2012GRMHDOutflows}.

\begin{figure*}
	\includegraphics[width=\textwidth]{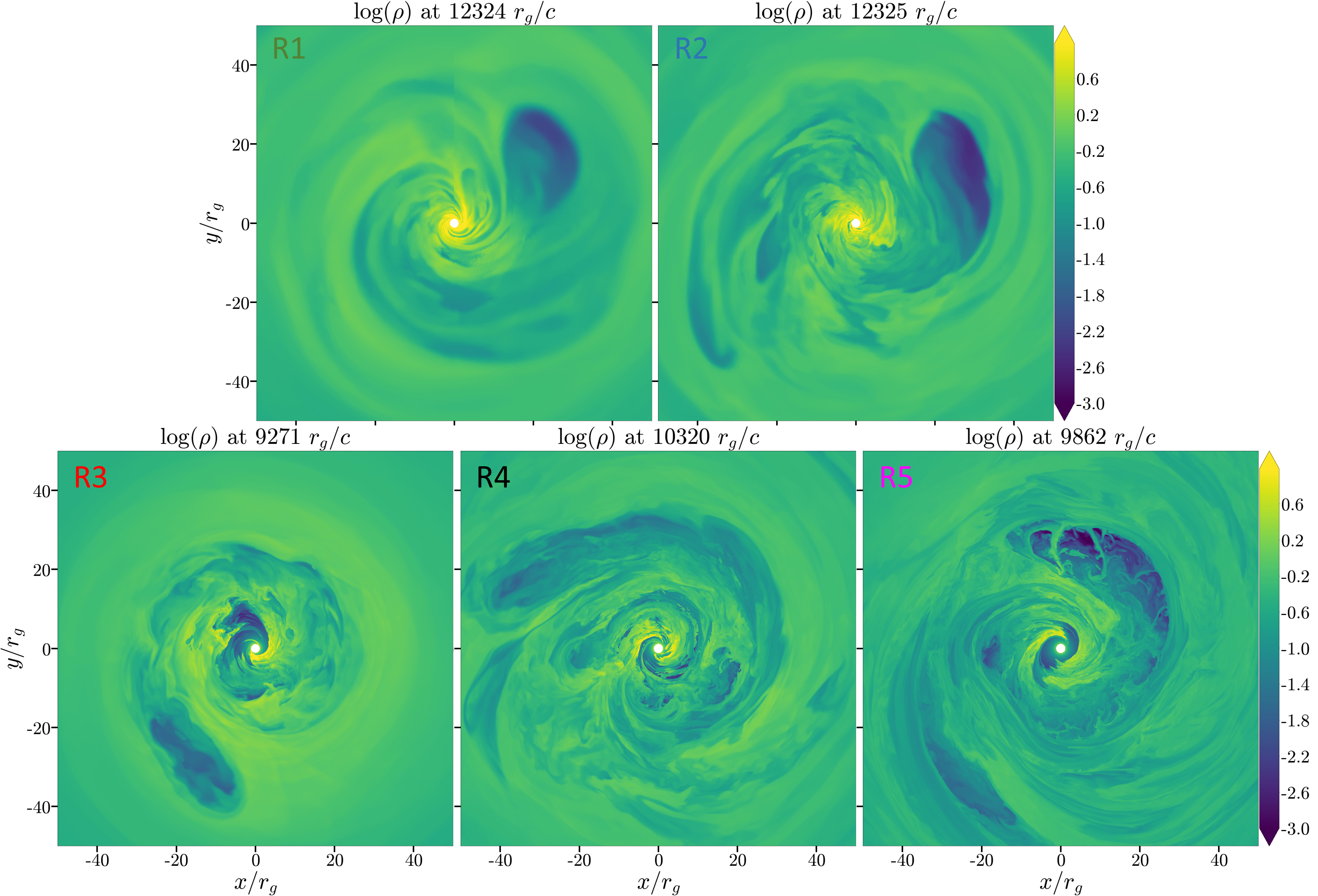}
    \caption{Equatorial 2D slices through $\theta=\pi/2$ of density $\rho$ showing low density and high magnetisation flux tubes  penetrating the accretion disk up to a radius of $\approx 40 r_g$. From left to right: resolution R1 to R2 (top) and R3 to R5 (bottom).}
    \label{fig:rhoxy}
\end{figure*}

Fig.~\ref{fig:rhoxy} displays the mid-plane density structures of the disks for all resolutions at different times. Low-density bubbles supported by magnetic pressure initially rise buoyantly away from the black hole and tend to orbit at a radius of $\approx 40 r_g$. The sizes of these bubbles are relatively constant regardless of resolution. Smaller flux eruptions, such as shorter ones, tend to yield smaller bubbles, while larger flux eruptions typically lead to the formation of larger bubbles. Gravity naturally pulls inward, directing denser material towards the center. In this case, the lower density bubbles are prone to gravitational mixing instabilities such as RTI \citep{Tchekhovskoy2011EfficientHole, Porth2021MNRAS.502.2023P, Ripperda2022BlackReconnection, Zhdankin2023PhRvR...5d3023Z}. At the lowest resolution R1, numerical diffusion causes the bubbles to mix. At higher resolution, finer structures are resolved because of mixing instabilities occurring at the surfaces of the bubbles.

There is a mixing of the low density bubbles with the higher density plasma in the accretion disk. This mixing causes mass to be advected inside the bubbles leading to a subsequent increase in the bubble density. At resolution R2, single plumes of high density plasma develop, and these plumes often split a bubble into two separate bubbles. From resolution R3 and higher, increasingly more plumes develop simultaneously in a single bubble. 

%This effect makes it more difficult to distinguish the bubbles from the surrounding material.

%\op{mention why this could be: increased numerical diffusion?}
%\op{how is this quantified? Not obvious as numerical diffusion can also be very efficient in mixing. }

Additional work is needed to quantitatively determine the resolution dependence on the time evolution of the volume, magnetisation, temperature and energy contained within the bubbles \citep[see e.g.][]{Porth2021MNRAS.502.2023P}, the exact role of the mixing instabilities, how mixing impacts the polarisation signatures of magnetic flux eruptions in the radio band \citep[e.g.][]{Jia2023MNRAS.526.2924J,Davelaar2023ApJ...959L...3D,Najafi-Ziyazi2024MNRAS.531.3961N}, or non-thermal particle acceleration \citep[e.g.][]{Zhdankin2023PhRvR...5d3023Z}

\subsubsection{Viscosity Parameters}
\label{sec:ViscosityParameters} % used for referring to this section from elsewhere

In this section, we study turbulence in the magnetised disk, which serves as a means to transfer angular momentum and energy. By making a decomposition of the velocity $\left \langle \vec{u}  \right \rangle _{\rho[\theta,\phi]}^{\textrm{disk}} + \delta \vec{u}$ (disk average plus fluctuation) in the momentum conservation equation, a turbulent stress tensor is obtained. This tensor has two components: Reynolds and Maxwell stress tensors, which represent the statistical correlations of velocity and magnetic field components respectively. Angular momentum transport and extraction of available energy from the flow to fuel the turbulent fluctuations is provided by the turbulent stress tensor in the orbiting plasma \citep{Balbus_Hawley_1998}. 

\begin{figure}
	\includegraphics[width=\columnwidth]{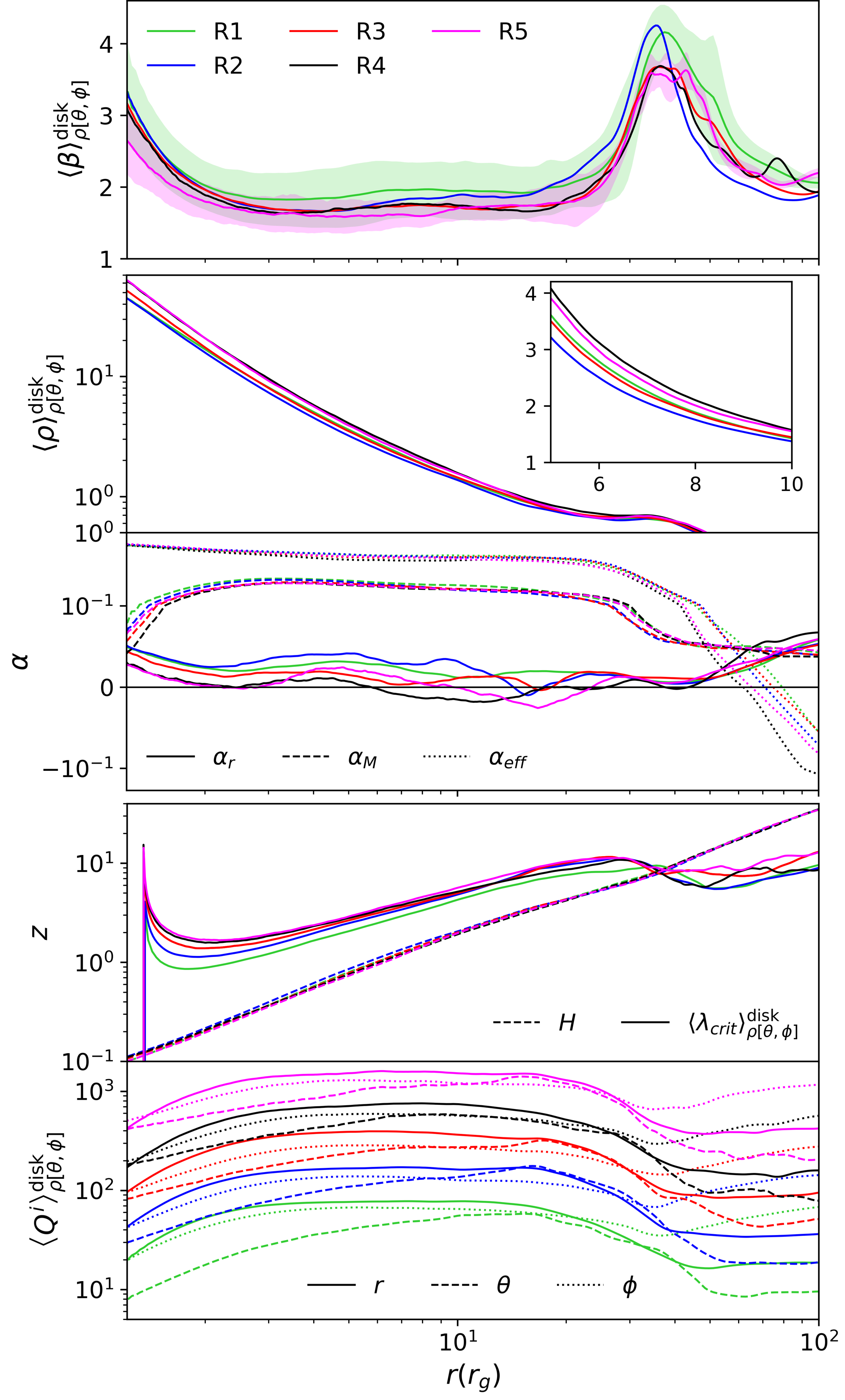}
    \caption{Time averaged, disk averaged profiles of variables as a function of radius. Fist panel: plasma beta $\beta=p_g/p_b$, gas-to-magnetic pressure ratio. Second panel: density. Third panel: viscosity parameters, Maxwell, Reynolds and Effective viscosity. Fourth panel: full height and critical wavelength. Fifth panel: quality factors. All the variables are averaged in time $t=[8-10]\times 10^{3} r_g/c$. For plasma beta, shaded region depicts the range of variation within one standard deviation for R1 and R5. In flux eruption events, highly magnetised plasma from the jet spine penetrates the disk mid-plane ($r\lesssim40r_g$) and propagates above and below the mid-plane ($40r_g \lesssim r \lesssim 100r_g$). As a consequence, the time averaged magnetic pressure (plasma beta) in the disk increases (decreases) with resolution.%\beta$ is convergent for R3 and higher.
    }
    \label{fig:DiskAvg}
\end{figure}

A viscosity parameter is often 
%\op{can be -- its often done but there are issues with this} 
related to the strength of the radial and azimuthal turbulent speed correlations, $\delta {u^r}=u^r - \left \langle  u^r \right \rangle_{\rho[\theta,\phi]}^{\textrm{disk}}$ and $\delta {u^{\phi}}=u^{\phi} - \left \langle  u^{\phi} \right \rangle_{\rho[\theta,\phi]}^{\textrm{disk}}$, respectively. We recast the turbulent stresses in the classical alpha-models as follows \citep[e.g.][]{shakura1973black,Balbus_Hawley_1998,Liska_Tchekhovskoy_Ingram2019_Bardeen-Petterson,ChatterjeeNarayan2022ApJ...941...30C}. The Reynolds viscosity is defined as:
%\op{We recast the turbulent stresses in the classical alpha-models as follows (some references here):}
\begin{equation}
    \alpha _r\equiv \frac{\left \langle \left ( \rho+u_g+p_g+b^2 \right )\delta {u^r}\sqrt{g_{rr}} \delta {u^{\phi}\sqrt{g_{\phi\phi}}} \right \rangle_{\rho[\theta,\phi]}^{\textrm{disk}}}{\left \langle p \right \rangle_{\rho[\theta,\phi]}^{\textrm{disk}}} \text{ ,}
	\label{eq:alphar}
\end{equation}
and the Maxwell viscosity as:
\begin{equation}
    \alpha _M\equiv -\frac{\left \langle b^r\sqrt{g_{rr}} b^{\phi}\sqrt{g_{\phi\phi}} \right \rangle_{\rho[\theta,\phi]}^{\textrm{disk}}}{\left \langle p \right \rangle_{\rho[\theta,\phi]}^{\textrm{disk}}}
	\label{eq:alphaM} \text{ .}
\end{equation}
%\op{This is a bit sloppy, I know many people just gloss over it but it trips up students starting into the field.  Can you add the sqrtgammarr etc which are actually in these equations? Nick Kaaz did an excellent job of defining these parameters.}
where, $p=b^2/2+p_g$ is the total pressure. The time averaged viscosities in the disk are presented in bottom left panel of Fig.~\ref{fig:DiskAvg}. We find that the Maxwell component dominates over the Reynolds component. The negative values of $\alpha _r$ indicate inward angular momentum transport and suggest convection-like behaviour \citep[e.g.][]{Begelman2022MNRAS.511.2040B}. The relatively low values of $\alpha _r$ indicate that the angular momentum transport due to turbulent convection is subdominant \citep[e.g.][]{ChatterjeeNarayan2022ApJ...941...30C}. As demonstrated in previous works \citep[e.g.][]{Liska_Tchekhovskoy_Ingram2019_Bardeen-Petterson}, the combination of both Reynolds and Maxwell viscosities is not sufficient to explain the total angular momentum transport, provided by the effective viscosity: 
\begin{equation}
    \alpha _{\textup{eff}}\equiv -\frac{\left \langle v^{r}\sqrt{g_{rr}} v^{\phi}\sqrt{g_{\phi\phi}} \right \rangle_{\rho[\theta,\phi]}^{\textrm{disk}}}{\left \langle c_s^2 \right \rangle_{\rho[\theta,\phi]}^{\textrm{disk}}}
	\label{eq:alphaeff} \text{ ,}
\end{equation}
%\op{see, here you added the hats... (but those are not defined, tststs...)}
where $v^r=u^r/u^t$ and $v^{\phi}=u^{\phi}/u^t$ are the physical radial and azimuthal velocity components, and the sound speed $c_s$ is:
\begin{equation}
    c_s\equiv \left | \frac{\gamma_{\mathrm{ad}} (\gamma_{\mathrm{ad}}-1)u_g}{\rho+u_g+p_g} \right |^{1/2} \text{ .}
	\label{eq:cs}
\end{equation}
%\op{check whether the adiabitic index is defined, I use $\hat{\gamma}$ for it. Or is it the thermal Lorentz factor?   Can you give a reference please? LS: Koushik2022 added in the beginning of the subsection}
The magnetorotational instability (MRI) is responsible for the generation of MHD turbulence that leads to enhanced outward angular momentum transport in accretion discs \citep{shakura1973black,Balbus_Hawley_1998, Pessah2006MNRAS.372..183P}. In this context, the Maxwell stress 
%\op{sorry many remarks on notation and terminology... I'd call it Maxwell stress. alpha is just a normalized version of the stress tensor component}
is associated to MRI. As seen in Fig.~\ref{fig:DiskAvg}, for $r=\left [ 3-50 \right ] r_g$, there is a significant contribution of MRI to transport angular momentum, since the Maxwell viscosity is only a factor of $\approx 3$ lower than the effective viscosity. For the very inner part of the accretion disk, $r<3 r_g$, $\alpha _{\textup{eff}}$ keeps growing while $\alpha _M$ drops significantly. For MAD models, magnetic ﬂux eruption–driven disk winds cause a strong vertical ﬂow of angular momentum \citep{ChatterjeeNarayan2022ApJ...941...30C} and jets can remove angular momentum close to the black hole \citep[e.g.][]{tchekhovskoy2012general,Narayan2022MNRAS.511.3795N}. 
%For a more rigorous treatment of angular momentum transport in GRMHD simulations (including the contribution of e.g. large-scale magnetic stresses, winds, compressible effects, nozzle shocks, disk tearing) see e.g. \citet{ChatterjeeNarayan2022ApJ...941...30C} for weakly magnetized (SANE) and MAD disks, \citet{Kaaz2023ApJ...955...72K} for warped thin disks and \citet{Das2022MNRAS.515.3144D} in the context of accreting neutron stars. %\op{also Kaaz and Koushik}. 

\subsubsection{Magnetorotational Instability}
\label{sec:MRI} % used for referring to this section from elsewhere

The MRI quality factor components $Q^i$ in the $i=[r,\theta,\phi]$-direction are defined as the number of cells available for resolving the fastest-growing MRI mode wavelength:
\begin{equation}
    Q^i\equiv \frac{2\pi r v_A^i}{\Delta x^iv} \text{ ,}
	\label{eq:Qi}
\end{equation}
where the size of the cell is $\Delta x^i \equiv dx^i\sqrt{g_{ii}}$. The Alfvén speed, which is the characteristic speed of transverse shear waves in a magnetized ﬂuid \citep{Balbus_Hawley_1998}, is defined as:
%The Alfvén velocity is the velocity of proton oscillation in response to a restoring force provided by an effective tension on the magnetic field lines in the plasma \op{that is not an explanation which is very common I think, I suggest to just leave it out. LS: Ok} :
\begin{equation}
    v_A^i\equiv \left ( \frac{\left | b^i b^ig_{ii} \right |}{\rho+u_g+p_g+b^2} \right )^{1/2} \text{ ,}
	\label{eq:vA}
\end{equation}
and the speed of the fluid is:
\begin{equation}
    v\equiv \frac{\left ( u^r u^r g_{rr} + u^{\theta} u^{\theta}g_{\theta\theta} + u^{\phi} u^{\phi}g_{\phi\phi}\right )^{1/2}}{u^t} \text{ .}
	\label{eq:Omega}
\end{equation}
%\op{umm, that is what you folks use? It should really be $u^\phi/u^t$ I think, but ok... LS: I agree $u^\phi/u^t$ makes more sense, but this is the definition in pp.py
%alf_speed = np.sqrt(np.abs(bu_proj[dir] * bu_proj[dir]) / (rho + bsq + (gam) * ug))
%vrot = np.sqrt((uu[3] * uu[3] * gcov[3][3] + uu[2] * uu[2] * gcov[2][2] + uu[1] * uu[1] * gcov[1][1])) / uu[0]
%wavelength = 2 * 3.14 * alf_speed * r / vrot
%if (dir == 1):
%   Q[dir] = wavelength / dx[dir] }

The quality factors in all three directions are presented in the bottom panel of Fig.~\ref{fig:DiskAvg}. %using the same conditions to select the disk as in Section~\ref{sec:Disk_averaged_profiles}. %the factors are calculated with two weighting conditions $\rho$ and $\rho b^2$. 
For all simulations, the quality factors are large enough to sustain the MRI-driven turbulence. The only exception is the polar component $Q^{\theta}$ both near the event horizon ($r<3r_g$) and further away in the disk ($40r_g \leq r \leq  100r_g$) for the lowest resolution R1. $Q^{\theta}<10$ due to the coarse size of the cell and to the relatively lower Alfvén speed in the polar direction. However, this does not significantly affect the time-averaged accretion rate or normalised magnetic flux, as they consistently converge across all five resolutions.

Apart from the saturation of magnetic flux, another characteristic of the MAD state is the suppression of the MRI due to the strong magnetic fields close to the the event horizon. \citet{Igumenshchev2003ApJ...592.1042I,Igumenshchev2008ApJ...677..317I} in pseudo-Newtonian MHD, and \citet{Tchekhovskoy2011EfficientHole,McKinney2012MNRAS.423.3083M} via GRMHD simulations, demonstrated the suppression of MRI when poloidal magnetic flux saturates near the BH, forming a MAD. %\citet{Begelman2022MNRAS.511.2040B} argue that MRI in the toroidal direction is not fully suppressed. 
\citet{MADresWhite} showed that the suppression of MRI is robust with resolution. From the general-relativistic dispersion relation connecting the Alfvén frequency (as seen by an observer at infinity) to oscillation frequency, \citet{Gammie2004TheMetric} proposed an instability criterion for the critical wavelength:
\begin{equation}
    \lambda_{crit} \equiv 2\pi v_A^z \frac{r^{3/2}+a}{3^{1/2}\left ( 1-2r^{-1}+a^2r^{-2} \right )^{1/2}} \text{ .}
	\label{eq:lambda_cr}
\end{equation}
The MRI is suppressed if the critical wavelength is bigger than the full height of the disk, $\left \langle \lambda_{crit} \right \rangle_{\rho[\theta,\phi]}^{\textrm{disk}}>H$ \citep{MADresWhite}. The full height is defined geometrically as $H\equiv r\left \langle \left | \theta - \pi/2 \right | \right \rangle_{\rho[\theta,\phi]}^{\textrm{disk}}$. %\op{did I miss it, what is R, which particular value of theta is chosen? Does this make sense? LS:Notation corrected}.

As a consequence of the disk magnetic field strength increasing with resolution, the MRI is more suppressed according to the criterion $\left \langle \lambda_{crit} \right \rangle_{\rho[\theta,\phi]}^{\textrm{disk}}>H$. The degree of the suppression of MRI in GRMHD simulation depends on whether the simulation is performed in 2D or 3D, alongside the numerical resolution used \citep[see e.g.][]{MADresWhite}. 
%\textcolor{magenta}{Add some references}. 
In axisymmetric 2D simulations, the MRI turbulence decays after time intervals of $\sim 1000 r_g/c$ \citep[e.g.][]{hide1982generalization,MRIBalbus,Guan2008ApJS..174..145G} and the most unstable axisymmetric linear mode of MRI
is stabilised \citep[e.g.][]{McKinney2012MNRAS.423.3083M,Begelman2022MNRAS.511.2040B}. \citet{Begelman2022MNRAS.511.2040B} found that MRI is not completely suppressed in non-axisymmetric MAD flows, where the saturation of magnetic flux is driven by radial convective/interchange instabilities triggered by a dominant toroidal field. The MRI is a MHD instability originally introduced for axisymmetric, magnetized accretion disks \citep{Balbus_Hawley_1998}. \citet{Goedbloed2022ApJS..259...65G} argue that axisymmetric
MRIs represent only a ﬁnite amount of unstable modes and that turbulence, accretion and dynamo activity in 3D disks is governed by the excitation of nonaxisymmetric super-Alfvénic rotational instabilities (SARIs).

%\begin{equation}
%    H/R_{\textup{thermal}}\equiv \frac{c_s u_t}{u^{\phi}}
%	\label{eq:H/R}
%\end{equation}

% % Example figure
% \begin{figure}
% 	% To include a figure from a file named example.*
% 	% Allowable file formats are eps or ps if compiling using latex
% 	% or pdf, png, jpg if compiling using pdflatex
% 	\includegraphics[width=\columnwidth]{Figures/Q_S_tavg.png}
%     \caption{Radial dependence of the quality factors in the disk. All averaged in time $t=[8-10]\times 10^{3} r_g/c$.}
%     \label{fig:QS}
% \end{figure}

%Since the MRI becomes more suppressed with increasing resolution \citep{MADresWhite}, the mass accretion rate is expected to decrease. 
%The suppression of MRI could also explain the enhancement of the disk density because angular momentum transport is less efficient through MHD turbulence (see the second row Fig.~\ref{fig:DiskAvg}). %The density accumulation in the mid-plane of the disk could be even higher when radiative cooling is included, affecting the dynamical evolution of the accreting plasma by reducing the gas pressure \citep{Yoon2020SpectralCooling}. 

\subsubsection{Plasmoids}
\label{sec:Plasmoids}

During an eruption in MAD states, the expulsion of magnetic flux goes through the magnetic reconnection of field lines in a current sheet in the equatorial plane. 
If this current sheet is sufficiently thin (see criteria for current sheet thickness $\delta$ below) such that plasmoid--mediated reconnection is resolved, then the current sheet can show signs of the tearing instability \citep{Ripperda2020ApJ...900..100R,Ripperda2022BlackReconnection}. Evidence of the plasmoid instability acting in the current sheet signifies that the rate at which magnetic energy is converted to heat, the reconnection rate, has converged to a value of $v_{\textup{rec}} = 0.01 v_A$ \citep{Bhattacharjee2009FastInstabilities,Uzdensky2010PhRvL.105w5002U}, where $v_A = \sqrt {\sigma / (\sigma+1)}c$ is the Alfvén speed. The plasmoid instability is typically triggered when the ratio between the alfvénic timescales and diffusion timescales, the Lundquist number $S\equiv v_A \omega / \eta \gtrsim 10^4$, where $\omega$ is the length of the current sheet \citep{Bhattacharjee2009FastInstabilities,Uzdensky2010PhRvL.105w5002U}. In this picture the thickness of the reconnection layer, $\delta$, or rather its aspect ratio $\delta / \omega$, must match the asymptotic reconnection rate to minimally resolve the tearing instability, $v_{\textup{rec}} =v_{in}/v_A = 0.01 = \delta/ \omega$, where $v_{in}$ is the inflow velocity into the layer, equal to the $\mathbf{E} \times \mathbf{B}/B^2$-velocity in the direction perpendicular to the layer. 

In ideal (GR)MHD, a current sheet, i.e., a magnetic null in the case of zero guide field as applicable to the equatorial current sheet in a MAD state \citep{Ripperda2022BlackReconnection}, is only captured by a single cell (since it is governed by a numerical resistivity, see below). Therefore, for a current sheet that tears at $r=2 r_g$ (approximately at the ergosphere), one requires 1 cell per $\delta=0.01 r_g$ for a current sheet of length $\omega=1 r_g$. Comparing to the domain size (poloidal angle $\theta \in [0,\pi]$), this results in $2 \pi r_g/ 0.01 r_g \gtrsim 600$ cells at $2 r_g$, implying that only simulations R4 and R5 have enough resolution in $\theta$ to minimally resolve the tearing instability near the horizon for current sheets of length $\omega \geq 1 r_g$, as indicated by our results below.

In ideal GRMHD simulations, such as those conducted here, reconnection is mediated via numerical resistivity $\eta _{\textup{num}}$ instead of an explicit resistivity $\eta$ \citep[][]{Ripperda2019ApJS..244...10R}. We can  determine whether our simulations are computed with sufficient numerical accuracy in order to resolve plasmoid-mediated reconnection, and thus form plasmoids by calculating the numerical Lundquist number $S\equiv v_A \omega/\eta _{\textup{num}}$, where $\eta _{\textup{num}} = \hat{C} \Delta x$ and the coefficient $\hat{C}$ depends on the accuracy and order of the reconstruction scheme \citep{Zhang2003PhysRevE.68.046709}.
%, simulations whether plasmoids form or not in the largest current sheets near the horizon. %(where the highest resolution is applied in a spherical grid). 
In the plasmoid instability regime, the rate of magnetic flux decay on the horizon converges to $\varphi_{BH}\propto e^{-t/500}$ \citep{Bransgrove2021PhRvL.127e5101B,Ripperda2022BlackReconnection}.
%\op{I'm convinced it scales like $\Delta x$, but I don't know the coefficient.  It depends on the accuracy of your reconstruction scheme and the order of it. It might well be quite a bit smaller than one.  You \textit{could} go so far as saying 'plasmoids are resolved at Ntheta=X, so we set the Lundquist nr to 10 000 there and obtain the coefficient this way. LS: I'll calculate the coefficient when I get the results of resolution R5.}
%\op{ugh that is grossly wrong, sorry!  Field lines can reconnect as soon as $S<\infty$, there is always numerical resistivity which makes sure this is the case also in an ideal mhd simulation. LS: sorry for the mistake}. 
In simulations R4 and R5 we observe plasmoids in current sheets of length $\omega \sim 10 r_g$ close to the event horizon, indicating that resolutions of a minimum of $2240 \times 1056 \times 1024$ are required to resolve the plasmoid instability, see Fig. 4 and Appendix C of \citet{Ripperda2022BlackReconnection}. Based on these results, the constant $\hat{C}\approx0.03$ is obtained by setting $10^4 = 10/\eta _{\textup{num,R4}}$, where $\eta _{\textup{num,R4}}$ is the numerical resistivity at $10r_g$ for R4. The Alfvén speed $v_A = \sqrt{25/26}c \approx c = 1$ for the reconnecting plasma because the reconnection is fed by the plasma in the jet at $\sigma _{\textup{max}}=25$. Due to the use of a spherical grid, it is harder to resolve plasmoid-mediated reconnection at larger radii and in smaller current sheets, due to the reduction in resolution at larger radii.

% The Lundquist number $S$ is used to determine if plasmoids are developed in a particular simulation, and defined as in \citet{Ripperda2022BlackReconnection}:
% \begin{equation}
%     S\equiv \frac{v_A \omega}{\eta _{\textup{num}}} \text{ ,}
% 	\label{eq:Si}
% \end{equation}

%The ratio of the Alfvén speed to the numerical resistivity as a function of radius is presented in the bottom panel of Fig.~\ref{fig:QS}. 

%\textcolor{red}{Note to Gibwa: check how the size of the cell is calculated in eq. 11 and 16. I'm using dx[i] = dxi / lowresi * np.sqrt(gcov[i, i, :, :, :, :])} LS: The set of simulations of this paper do not use AMR

%\begin{figure}
%   \includegraphics[width=\columnwidth]{Figures/LundquistN50_5-10.png}
%    \caption{Ludquist numbers as a function of radius and time.}
%    \label{fig:S}
%\end{figure}

\subsubsection{Disk profiles}
\label{sec:Disk_averaged_profiles}

The time averaged profiles of the disk plasma beta (ratio of gas-to-magnetic pressure $\beta=p_g/p_b$), density, viscosity parameters (Maxwell, Reynolds and effective viscosity), full height, critical wavelength and quality factors are presented in Fig.~\ref{fig:DiskAvg}. All variables presented in Fig.~\ref{fig:DiskAvg} are averaged between $8$ to $10\times 10^{3} r_g/c$.

\begin{figure*}
	\includegraphics[width=\textwidth]{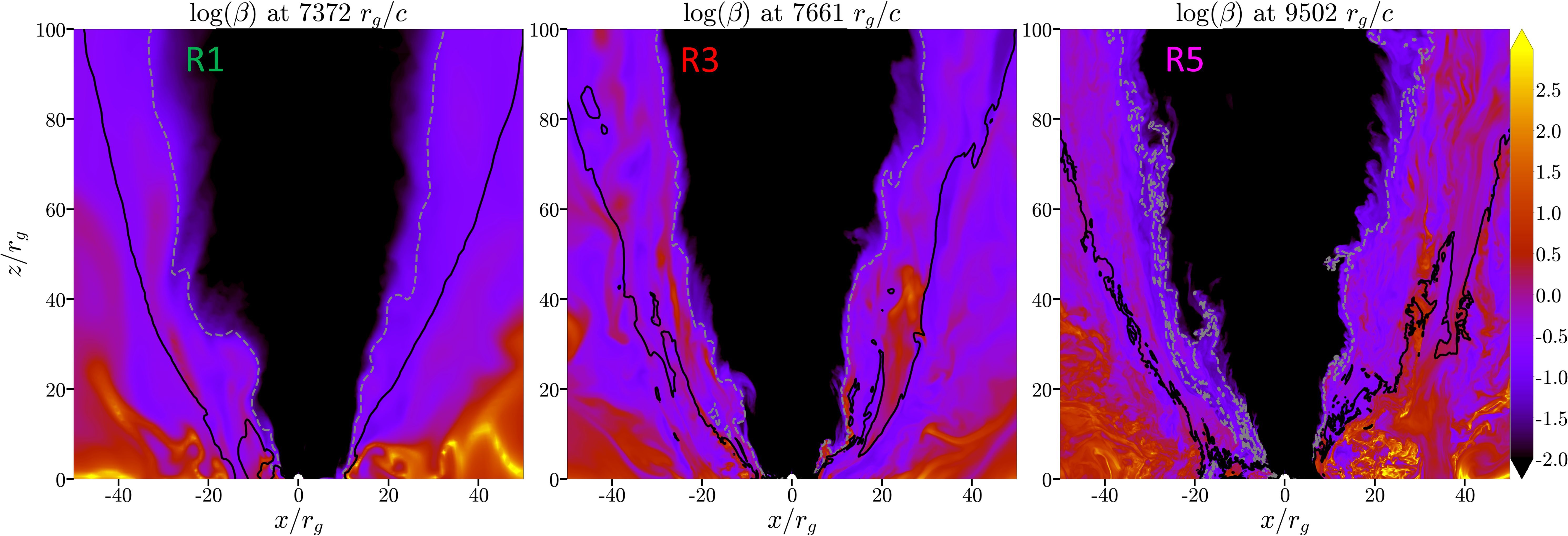}
    \caption{Meridian 2D slices through $\phi=0$ for plasma beta $\beta=p_g/p_b$. From left to right: resolution R1, R3 and R5. The grey dashed and black lines correspond to surfaces where $\sigma = 1$ and $Be=1.02$, respectively. We partially cover the spine with a zero screen activated for $\sigma > 3$, allowing visibility of plasma within the range of $1 < \sigma < 3$. As resolution increases, the polar axis features a smaller volume and therefore there is less dissipation of magnetic energy along the pole. In flux eruption events, highly magnetised plasma from the jet spine penetrates the disk, causing the time averaged plasma beta in the disk to decrease with resolution.}
    %\op{what are the curves here?}}
    \label{fig:beta}
\end{figure*}

2D slices through azimuthal angle $\phi=0$ of plasma beta $\beta$ are shown in Fig.~\ref{fig:beta}. We find that the jet spine gets more magnetised with increasing resolution as demonstrated in Section~\ref{sec:Jet} (Fig.~\ref{fig:Jet-spine}). During flux eruption events, highly magnetized flux tubes originating from the jet spine penetrate the disk and propagate until $r\lesssim40r_g$. The pressure maximum of the initialised Fishbone-Moncrief torus is located at $r=41 r_g$-- at around this radius, the flux tubes split into two and are ejected as outflows above and below the disk mid-plane. These ejections increase the time-averaged disk magnetic pressure, and consequently reduce the plasma beta in the disk, as observed in Fig.~\ref{fig:DiskAvg}. Plasma beta in the disk for $r\gtrsim3r_g$ is convergent for resolutions R3 and higher. For resolution R5, plasma beta for $r\lesssim3r_g$ is slightly lower than the rest of simulations.  Very close to the BH $r\lesssim 3r_g$, the Maxwell viscosity drops significantly for all resolutions. In this inner part of the accretion disk, the vertical magnetic pressure is responsible for launching winds that transport angular momentum outwards \citep[e.g.][]{ChatterjeeNarayan2022ApJ...941...30C}.
%The temperature in the funnel can reach values up to $\sim 10^{13} K$, but it typically produces negligible emission as a consequence of the extremely low densities in this region. \textcolor{magenta}{note floors...}

%\citet{Begelman2022MNRAS.511.2040B} found that MRI is not completely suppressed in non-axisymmetric MAD flows, where the saturation of magnetic flux is driven by radial convective/interchange instabilities triggered by a dominant toroidal field. Very close to the BH $r\lesssim 3r_g$, the Maxwell viscosity drops significantly for all resolutions (Fig.~\ref{fig:DiskAvg}). In this inner part of the accretion disk, the vertical magnetic pressure is responsible for launching winds that transport angular momentum outwards \citep[e.g.][]{ChatterjeeNarayan2022ApJ...941...30C}. %For $3r_g\lesssim r \lesssim 20r_g$ the magnetic pressure supports the accretion disk against gravity during the magnetic flux eruptions. This cause the low density bubbles to be unstable to the RTI \citep[e.g.][]{McKinney2012MNRAS.423.3083M}.

%\op{interesting, in principle the MRI is not the only thing that can act, you could link up with the discussion by Begelman et al.  (Rayleigh-Taylor and convective-type instabilities are on the market too).}

%% file: 4Jet/Jet.tex
\subsection{Jet Properties}
\label{sec:Jet}

The averaged profile of a jet variable does not involve any density or magnetisation weight. For the jet spine:
\begin{equation}
    \left \langle X \right \rangle_{[\theta,\phi]}^{\textrm{spine}} = \frac{\iint X \left ( \sigma > 1 \right ) \sqrt{-g}d\theta d\phi}{\iint \left ( \sigma > 1 \right ) \sqrt{-g}d\theta d\phi} \text{ ,}
	\label{eq:AverageFunnel}
\end{equation}
and for the jet sheath:
\begin{equation}
    \left \langle X \right \rangle_{[\theta,\phi]}^{\textrm{sheath}} = \frac{\iint_{\left | \pi-\theta \right |\leqslant \pi/6} X \left ( \sigma < 1 \right ) \left ( Be > 1.02 \right )\sqrt{-g}d\theta d\phi}{\iint_{\left | \pi-\theta \right |\leqslant \pi/6} \left ( \sigma < 1 \right ) \left ( Be > 1.02 \right )\sqrt{-g}d\theta d\phi} \text{ .}
	\label{eq:AverageSheath}
\end{equation}

We follow the methodology used in \citet{JetsKoushik} to investigate the instabilities occurring in the jet sheath-spine interface. Decomposing the stress-energy tensor ${T^{r}}_{t}\equiv{\left ( T_H \right )^{r}}_{t}+{\left ( T_M \right )^{r}}_{t}$ (Equation~\ref{eq:Tud}), with the hydro component:
\begin{equation}
    {\left ( T_H \right )^{r}}_{t}\equiv \left (\rho+u_g+p_g \right ) u^{r} u_{t} \text{ ,}
	\label{eq:T_Hrt}
\end{equation}
and magnetic component:
\begin{equation}
    {\left ( T_M \right )^{r}}_{t}\equiv b^2 u^{r} u_{t}-b^{r} b_{t} \text{ ,}
	\label{eq:T_Mrt}
\end{equation}

We define the total specific energy, measuring the maximum Lorentz factor if all forms of energy are converted into kinetic energy,
\begin{equation}
    \mu\equiv \frac{\left \langle {T^{r}}_{t} \right \rangle_{[\theta,\phi]}^{\textrm{spine}}}{\left \langle -\rho u^r \right \rangle_{[\theta,\phi]}^{\textrm{spine}}} \text{ ,}
	\label{eq:mu}
\end{equation}
the magnetisation, as a measure of conversion efficiency of kinetic to magnetic energy,
\begin{equation}
    \hat{\sigma}\equiv \frac{\left \langle {\left ( T_M \right )^{r}}_{t} \right \rangle_{[\theta,\phi]}^{\textrm{spine}}}{\left \langle {\left ( T_H \right )^{r}}_{t} \right \rangle_{[\theta,\phi]}^{\textrm{spine}}}
	\label{eq:sigma} \text{ ,}
\end{equation}
the specific enthalpy,
\begin{equation}
    h\equiv \frac{\left \langle {T^{r}}_{t}\left(u_g + p_g  \right )/\rho \right \rangle_{[\theta,\phi]}^{\textrm{spine}}}{\left \langle {T^{r}}_{t} \right \rangle_{[\theta,\phi]}^{\textrm{spine}}} \text{ ,}
	\label{eq:h}
\end{equation}
and the Lorentz factor,
\begin{equation}
    \gamma\equiv \frac{\left \langle {T^{r}}_{t}u^t\sqrt{-1/g^{tt}} \right \rangle_{[\theta,\phi]}^{\textrm{spine}}}{\left \langle {T^{r}}_{t} \right \rangle_{[\theta,\phi]}^{\textrm{spine}}} \text{ .}
	\label{eq:gamma}
\end{equation}

The energy equation can be obtained by combining the four previously defined variables:
\begin{equation}
    \mu = \gamma\left ( \hat{\sigma} +1\right )\left ( h + 1 \right ) \text{ .}
	\label{eq:Eeq}
\end{equation}

%All four variables ($\mu$, $\gamma$, $\sigma$ and $h$) are calculated for the up and down jet funnel as a function of distance from the poles of the black hole, by averaging in $\theta$ and $\phi$. 

\subsubsection{Jet spine}

\begin{figure}
	\includegraphics[width=\columnwidth]{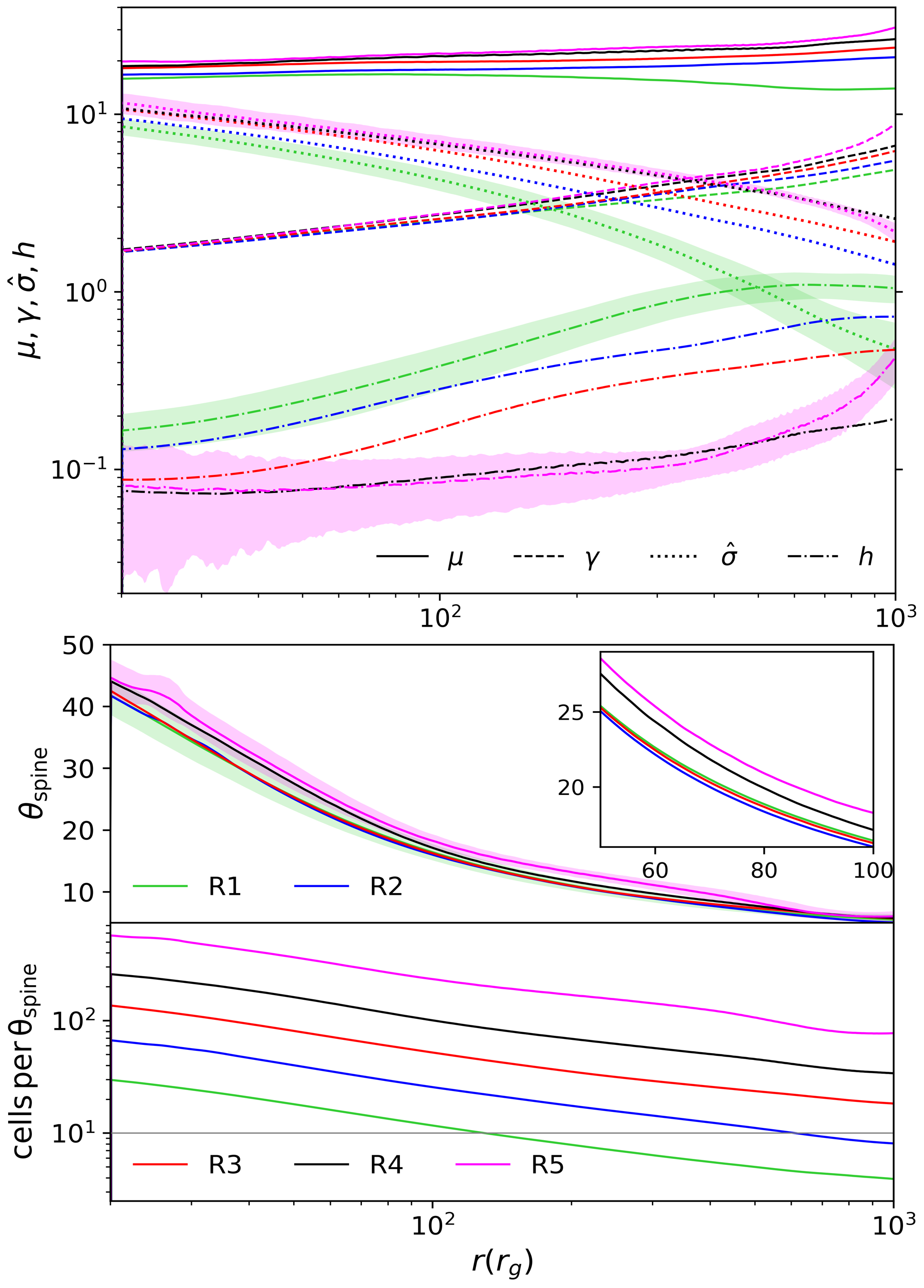}
    \caption{Jet spine ($\sigma>1$) properties as a function of $r$, averaged in time $t=[8-10]\times 10^{3} r_g/c$. Total specific energy, Lorentz factor, magnetisation and enthalpy (top panel). Half opening angle (middle panel) and cells per half opening angle (bottom panel). Shaded regions depict the range of variation within one standard deviation for R1 and R5. $\hat{\sigma}$ and $h$ are not physically trustworthy. Additionally, the jet spine is wider for R4 and R5.}
    \label{fig:Jet-spine}
\end{figure}

The top panel of Fig.~\ref{fig:Jet-spine} shows the total specific energy, magnetisation, enthalpy, Lorentz factor and half opening angle of the upper jet spine (Equation~\ref{eq:AverageFunnel}), each of which are averaged over the time interval $t=[8-10]\times 10^{3} r_g/c$. We find that the Lorentz factor is nearly the same for all resolutions presented, and increases from approximately $2$ to $7$ with radius $20 - 10^3r_g$. The total specific energy is convergent and nearly constant with radius. However, the magnetisation and enthalpy are convergent only for resolutions R4 and R5 and are not physically trustworthy.

Since numerical dissipation is proportional to the cell size, as the resolution increases the cells get smaller. Consequently, cells near the polar axis have a smaller volume which results in less dissipation of magnetic energy and heating along the pole. As a result the jet spine gets more magnetised, $\hat{\sigma}$ drops notably slower with radius, and the internal gas energy and pressure decrease with resolution as shown in Fig.~\ref{fig:rho512}. Properties such as magnetisation and enthalpy in the jet become unreliable if number of cells per half opening angle drops below 10 (see bottom panel of Fig.~\ref{fig:Jet-spine}). Additionally, the polar axis introduces considerable uncertainty into thermodynamics-- while the transmissive boundary conditions used at the poles effectively capture the flow of energy, they artificially change the conversion of one energy source into another.
%This can be seen as well in Fig.~\ref{fig:rho512}. 

%\op{reference once more how these regions are defined.  How regions are masked plays a role in these energy profiles, so the reader wonders about that.}

%For the specific case of M87 extended jet
%with a rectangular grid of $261\times261\times729$%, \citet{Nakamura2004PoyntingStructures,Nakamura2018ParabolicM87} argue that a growing kink instability associated with MHD shocks in the highly magnetized relativistic jet may be responsible for organizing the conical jet in M87 at the kilo-parsec scale.  In \citet{Tchekhovskoy2016Three-dimensionalDichotomy}, kink instabilities control the jet morphology and can lead to the Fanaroff–Riley dichotomy of AGN jets, and also constitute the main dissipation mechanism responsible for powering Gamma Ray Burst (GRB) prompt emission (\citet{Bromberg2016RelativisticDissipation}). An important factor determining brightness of individual flares may be related to the nonaxisymmetry of the kink instability \citet{Sikora2005AreFlux}.

%\begin{figure}
%	\includegraphics[width=\columnwidth]{Figures/Jet-sheath_variables_tavg.png}
%    \caption{Jet-sheath properties as a function of $z$, averaged in time $t=[5-10]\times 10^{3} r_g/c$}
%    \label{fig:Jet-sheath}
%\end{figure}

%\op{First point out what's different (also give some numbers for Lorentz factor); There is significantly less heating of jet material in high-res and sigma drops notably slower.  This looks like magnetic field is dissipated efficiently at low res.  How does this connect to the discussion in the beginning about instabilities?  One would hope that once instabilities are resolved in the non-linear stage, the heating and mixing rate is converged.  This seems not to be the case here.  Quite the opposite: higher resolutions (better ability to resolve instabilities) have lower dissipation rate.  In the end, are these results maybe due to the way in which you mask?  Discuss this in the context of Koushik's 2019 paper (e.g. his Section 4.1.3).  Where are the pinch instabilities?}

The half opening angle of the jet spine is slightly higher for R4 and R5, especially close to the base of the jet. The standard deviation of the half opening angle decreases with radius. Close to the black hole, $r<50r_g$, we see high fluctuations in the width of the jet spine and sheath due to the magnetic flux eruptions in the MAD states.

\subsubsection{Jet sheath}

\begin{figure}
	\includegraphics[width=\columnwidth]{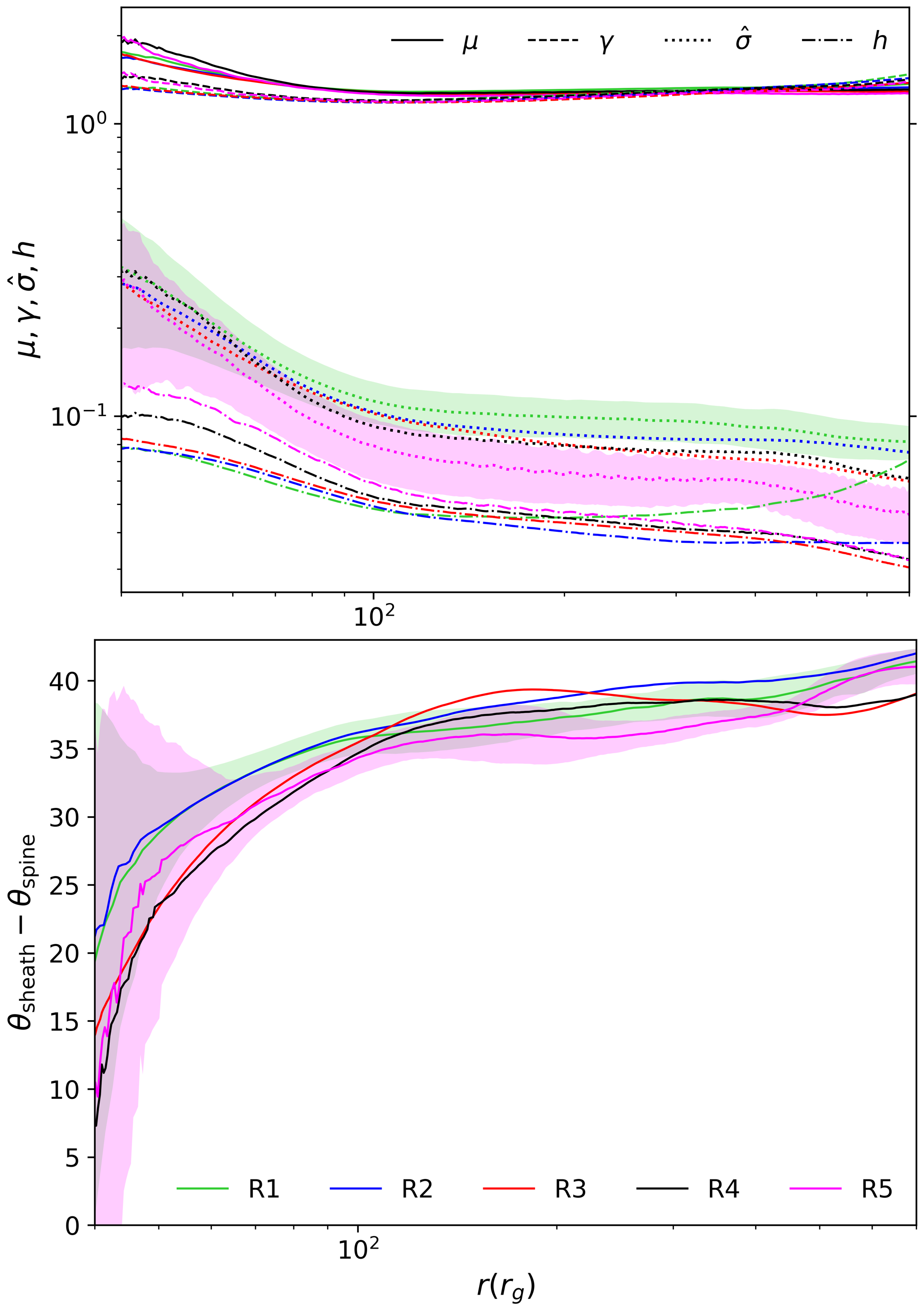}
    \caption{Jet sheath properties as a function of $r$, averaged in time $t=[8-10]\times 10^{3} r_g/c$. Total specific energy, Lorentz factor, magnetisation and enthalpy (top panel). Half opening angle of the jet sheath measured from the jet spine (bottom panel). Shaded regions depict the range of variation within one standard deviation for R1 and R5. Mixing instabilities occurring in the jet sheath-spine interface cause a reduction of magnetisation and increase of enthalpy with resolution.}
    \label{fig:Jet-sheath}
\end{figure}

The top panel of Fig.~\ref{fig:Jet-sheath} shows the total specific energy, magnetisation, enthalpy, Lorentz factor, and half opening angle of the jet sheath, all averaged over the time interval $t=[8-10]\times 10^{3} r_g/c$. In general, all these variables in the sheath are lower than the corresponding values in the spine, except the enthalpy for $r\lesssim60r_g$ where $h_{\mathrm{sheath}}>h_{\mathrm{spine}}$ for R4 and R5. The enthalpy for R1 has an inflection point around $100r_g$, where $h$ starts to increase due to enhanced numerical dissipation caused by a lack of resolution in the polar direction.

\begin{figure}
	\includegraphics[width=\columnwidth]{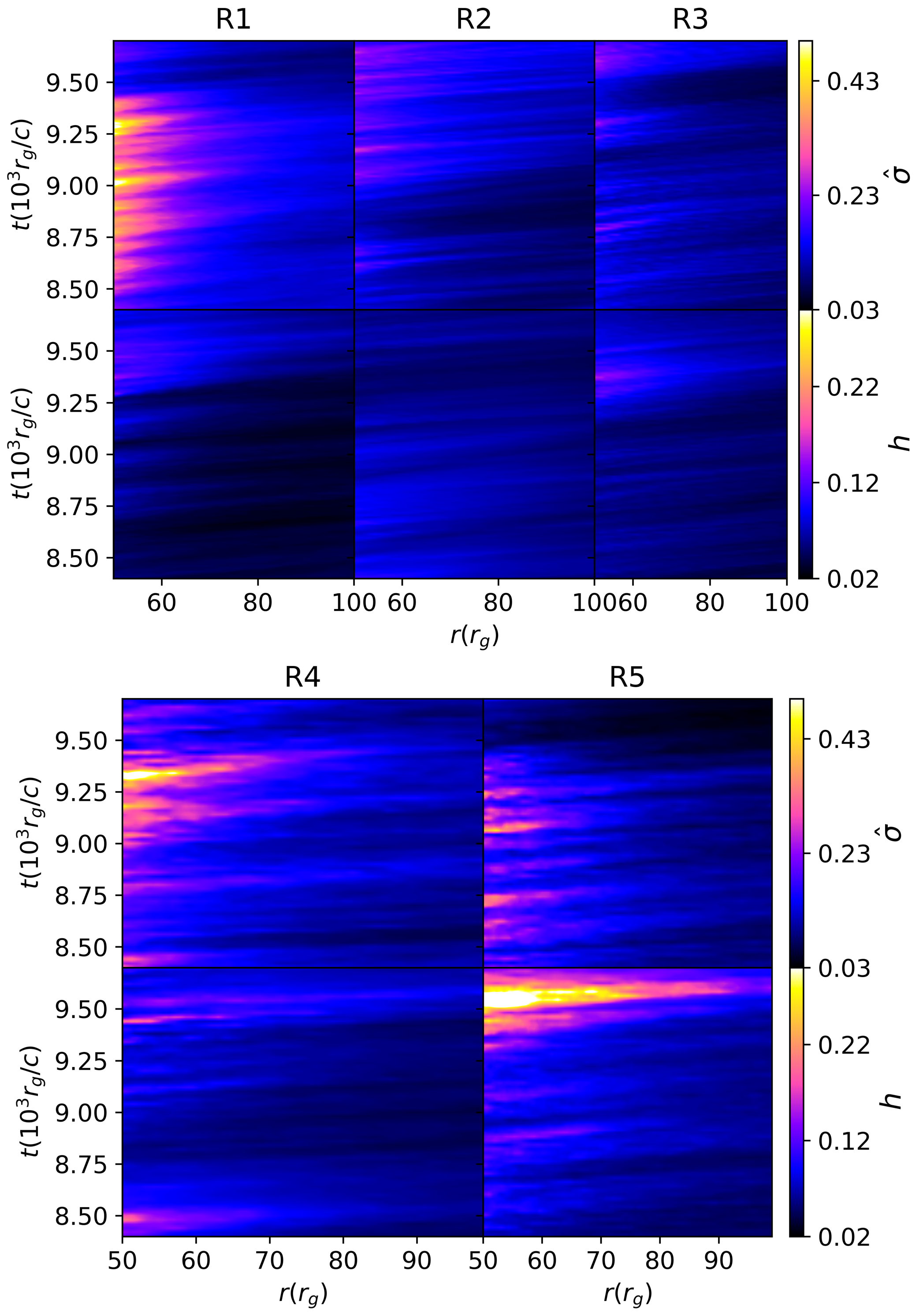}
    \caption{Space-time diagrams of magnetisation (top panels) and enthalpy (bottom panels) of the jet sheath. From left to right: resolution R1 to R5. A localised decrease in $\hat{\sigma}$ is accompanied by a simultaneous rise in enthalpy, generating a wave that propagates outwards.}
    \label{fig:sigma_h}
\end{figure}

When a flux eruption occurs, $\hat{\sigma}$ drops locally at the base of the jet generating a wave that propagates outwards. This wave is accompanied by a simultaneous increase of enthalpy, creating a thermal pressure gradient that slows down the jet outﬂow (see Fig.~\ref{fig:sigma_h}). However, the Lorentz factor of the jet $\gamma$ only decreases a few percent for all resolutions. In contrast, \citet{JetsKoushik} found a significant reduction of $\gamma$ at $r\approx10^3 r_g$ when pinch instabilities cause magnetic dissipation, that raise the jet specific enthalpy to order unity. The fact that we do not see such a significant reduction of $\gamma$ in our simulations is a consequence of the much smaller disc that collimates the jet out to approximately $500r_g$. For $r>500 r_g$, the diminishing confining pressure exerted by the disk wind triggers a lateral expansion of the jet, effectively suppressing pinch instabilities \citep[see, e.g.,][]{Porth2015MNRAS.452.1089P}. This expansion of the jet leads to acceleration and increase of $\gamma$.

Mixing instabilities due to shearing motions across the interface between the jet and the accretion disk can potentially cause the dissipation of magnetic energy into heat. Resolving these mixing instabilities is crucial for comprehending the energetic dynamics of jets at large radii, as mixing will lead to mass loading of the jet. Close to BH, the jet is primarily loaded by material from density floors. We see wave-like features in the jet sheath, which become more prominent (less diffusive) at higher resolutions. These wave-like features may induce mixing between jet, sheath and disk (see Fig.~\ref{fig:rho512} and \ref{fig:beta}). In addition to affecting the degree of mass loading in the jet, the wave-like features in the jet sheath may also affect the magnetic field in the proximity of the jet sheath: \citet{Davelaar2023ApJ...959L...3D} found that waves propagating along the jet-wind shear layer alter the orientation of the magnetic field lines.

For $r\lesssim100r_g$, the half opening angle of the sheath features strong fluctuations and on average the sheath is thinner at resolutions R3, R4 and R5 (see bottom panel of Fig.~\ref{fig:Jet-sheath}). Between $100$ and $\approx500r_g$, the sheath is thinner at resolution R5. This slimming of the jet sheath can result in an increase of enthalpy and reduction of magnetisation in the sheath, while keeping an approximately constant total specific energy for all resolutions: These results have a direct impact on the temperature, as described in Section~\ref{sec:TpTe}.%\citet{Davelaar2023ApJ...959L...3D} found that waves propagating along the jet-wind shear layer alter the orientation of the magnetic field lines. %and that causes a drop of the linear polarisation fraction of the observed synchrotron emission.

\subsubsection{Proton temperature}
\label{sec:TpTe}

%In computing the average values of the proton and electron temperature, we neglect the jet spine region, where the density and magnetization is set by numerical floor values.

\begin{figure}
	\includegraphics[width=\columnwidth]{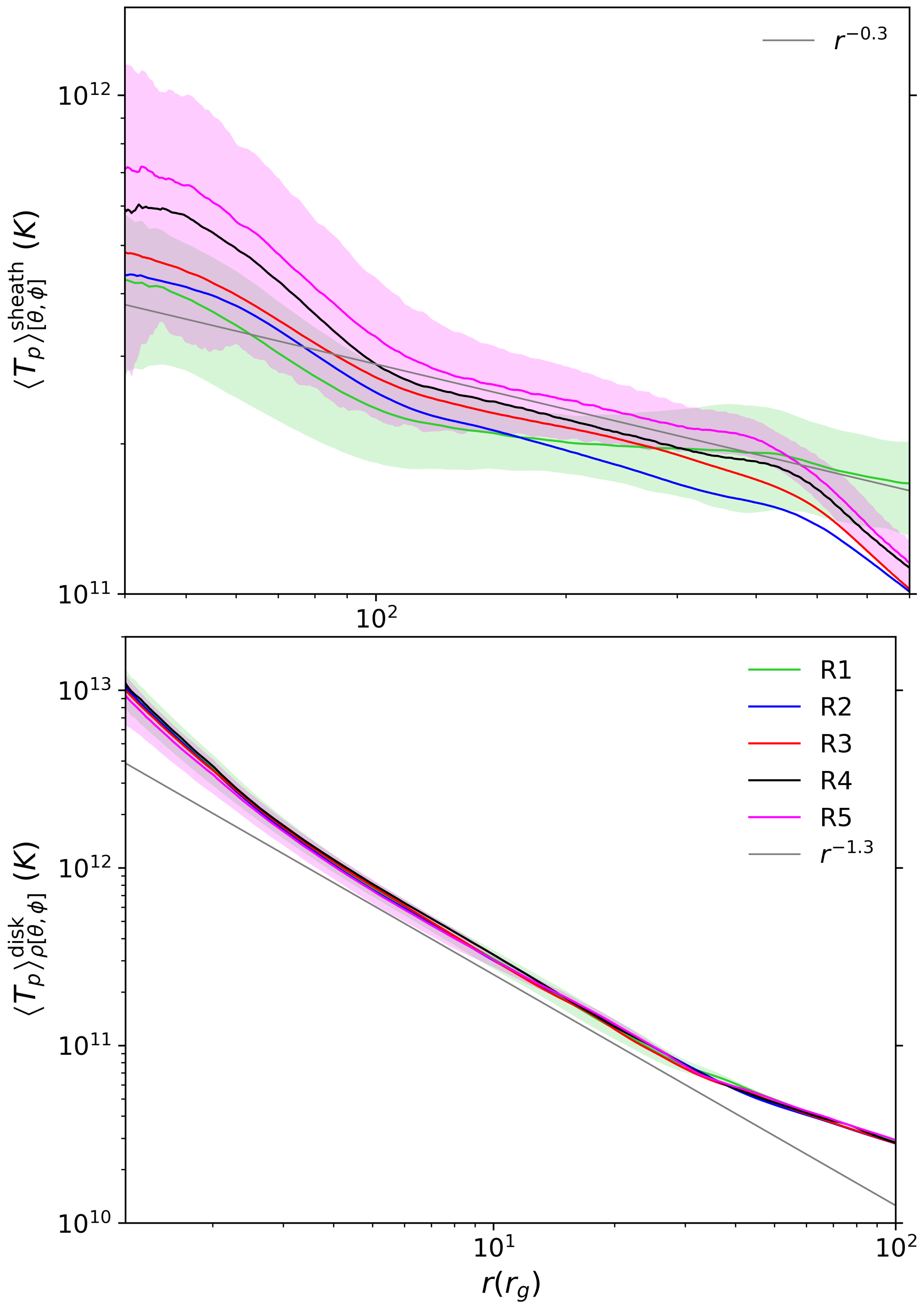}
    \caption{Proton temperature in the jet sheath (top panel) and in the disk (bottom panel). All averaged in $t=[8-10]\times 10^{3} r_g/c$. Shaded regions depict the range of variation within one standard deviation for R1 and R5.}
    \label{fig:Tp}
\end{figure}

The proton temperature in the jet sheath and disk is presented in Fig.~\ref{fig:Tp}. The protons in the disk are near virial temperature $\propto  r^{-1.3}$, and $T_p$ in the disk is converged for all resolutions we consider. On the other hand, $T_p$ in the sheath is found to increase with increasing resolution. %The heating in the jet sheath is enhanced as physical processes are better resolved in higher resolution. 
%\op{I think you should stress that jet-sheath (proton) temperatures are non convergent a bit more as one of the takeaways here (maybe move the discussion to the end of the paragraph).  }
Processes causing a mixing of the colder jet sheath with the hotter jet spine can include: (1) mixing instabilities due to shearing motion across the interface between the jet sheath and funnel, (2) waves generated at the jet base in MAD states that propagate along the jets and, (3) wobbling of the jet itself \citep[see e.g.][]{Wong2021ApJ...914...55W,Davelaar2023ApJ...959L...3D}. In addition, the increase we see in the magnetisation of the jet spine as the resolution is increased also impacts $T_p$ in the sheath. Highly magnetised plasma from the jet spine supplies matter to the current sheet, with $T_p \propto \sigma_{spine}$, and the reconnection exhaust deposits this hot plasma in the jet sheath up to at least $20r_g$ for resolutions R4 and R5 \citep{Ripperda2022BlackReconnection}. For $r<100r_g$, the sheath is thinner at higher resolution (see Fig.~\ref{fig:Jet-sheath}), the amount of dissipation/heating gets distributed across a smaller volume of plasma, leading to higher maximum temperature. The jet sheath (proton) temperatures are non fully convergent with resolution. Despite this, $T_p$ varies by less than a factor of $\sim2$ between the lowest and highest resolutions.
%\op{there is also an effect that plasmoids from reconnection of the equatorial current sheet are expelled along the jet sheath.  Bart knows this, maybe it also plays a role here?}
%\op{in which way, but which mechanism?}

%As seen previously in Fig.~\ref{fig:DiskAvg}, $\beta$ in the disk decreases with resolution, however, there is no significant difference between $T_e$ in the disk. This is a consequence of the parameterized coupling of the electron and proton temperatures for one particular choice of $R_{\textup{high}}$. 

The proton temperature in the jet sheath follows a shallower profile, $\propto  r^{-0.3}$, than in the disk $\propto  r^{-1.3}$. The jet is well collimated by the large disk when the numerical resolution is adequate for capturing the jet at large radii away from the black hole. The proton temperature profile in the jet sheath is different for the lowest resolution level R1. In this case, $T_p$ remains nearly constant for radii between 100 and 500 $r_g$, and then follows a power law of $\propto r^{-0.3}$ at larger radii. This discrepancy arises from a lack of resolution in the polar direction at $r>100r_g$ along the jet axis for R1 compared to the other resolutions we present.

Our simulations show that the adiabatic decompression of the jet sheath begins at around $500 r_g$, where the temperature gradient is much steeper than $r^{-0.3}$ (Fig.~\ref{fig:Tp}). It is worth mentioning that the change in steepness of the temperature gradient depends on the size of the initial torus, and it is unknown whether this adiabatic decompression happens in real systems. For instance, observations utilising the Very Large Array (VLA) have revealed a parabolic collimation profile in the \M87 jet extending up to $10^5 r_g$ \citep{AsadaNakamura2012ApJ...745L..28A,Hada2013ApJ...775...70H,Mertens2016A&A...595A..54M,Kim2018A&A...616A.188K,Nakamura2018ApJ...868..146N}.

%% file: 6Conclusions/Conclusions.tex
\section{Conclusions}
\label{sec:Conclusions}

GRMHD codes face significant numerical challenges when used to perform simulations in the MAD regime. The higher magnetic flux characteristic of the MAD regime leads to new dynamics, including interchange-type accretion modes, suppression of MRI and flux eruptions governed by plasmoid dominated magnetic reconnection. The occurrence and dynamics of these phenomena may be affected by the numerical resolution of GRMHD MAD simulations. Recently, \citet{MADresWhite} conducted a resolution study of MAD accretion flows, with simulations reaching a maximum effective resolution of $512 \times 256 \times 512$. %achieved with four different levels of Static Mesh Refinement and with two different spatial reconstruction algorithms. %\citep{white2016extension}
They showed that the general large-scale structure of the accretion flow is robust with resolution, for the resolutions they considered.
%and PPM was proven to better capture spatial variability. 
However, they found that the spatial structure and Lorentz factor of the jet, small-scale features of the turbulence and variability of modelled synchrotron emission were not fully converged with resolution.

We have conducted a study of MAD states and jets across a wider range of resolutions than has been previously explored, up to resolution of $5375 \times 2304 \times 2304$ in a logarithmic spherical-polar grid, using existing simulations in \citet{Ripperda2022BlackReconnection}. Such extreme resolution is needed to achieve convergence in the reconnection rate in the plasmoid-dominated regime in GRMHD, which is important for the timescale of variability. We divided the system in three components: the jet spine (relativistic and strongly magnetised outflow), the jet sheath (mildly relativistic outflow) and the disk. Below, we elucidate how our findings might influence ongoing studies within the black hole accretion community.

%\subsubsection{Viscosity and MRI}

We find that the time-averaged disk properties are consistent for all resolutions with only small differences in flux variability, plasma beta and inferred viscosity parameters close to the BH. %The Reynolds viscosity is determined indirectly through statistical correlations of the turbulent velocity components. %The angular momentum transport %of MADs is dominated by magnetic driven outflows, 
%attributed to the Maxwell viscosity (due to MRI) represents a third of the effective viscosity. 
For resolution $2240 \times 1056 \times 1024$ and higher, we find that the time-averaged Reynolds viscosity, attributed to turbulent convection, indicates inward angular momentum transport around $10r_g$.
%For resolution $2240 \times 1056 \times 1024$ and higher, we find that time-averaged Reynolds viscosity is negative around $10r_g$, indicating an inward angular momentum transport attributed to turbulent convection. %During flux eruptions, at the lowest resolution $288 \times 128 \times 128$, numerical diffusion causes flux tubes to mix with the higher density plasma in the accretion disk. 
Higher resolutions resolve finer structures when mixing instabilities occur at the surfaces of flux tubes during flux eruptions (see Fig.~\ref{fig:rhoxy}). This mixing may function as mechanism for energy dissipation, non-thermal particle acceleration and flare generation across various wavelengths.
%Resolution R1 is below the MRI resolving criteria outlined by \citet{hawley2011assessing} ($Q^z\geqslant 10$ and $Q^{\phi}\geqslant 20-25$).
%However, the Reynolds viscosity contribution is subdominant and does not substantially affect the accretion rates.

%\subsection{Jet properties}
%\subsubsection{Jet funnel}

For the jet spine, the total specific energy and Lorentz factor is nearly the same for all resolutions we consider. We find that the magnetisation and enthalpy are convergent for resolution $2240 \times 1056 \times 1024$ and higher. However, these four jet spine properties do depend on, and are governed by, the chosen magnetisation ceiling in the code setup. For this reason, we recommend caution in drawing conclusions based on jet dynamics in general, as the jet spine is dominated by numerical floors and the magnetisation and enthalpy are not physically reliable. %, though small differences arise for radii greater than $100r_g$. 
%The half opening angle of the jet spine is few degrees larger for resolutions $2240 \times 1056 \times 1024$ and $5376 \times 2304 \times 2304$. 

%The magnetisation and enthalpy in the jet spine are convergent for resolution R4 and R5. As the resolution increases, the polar axis features a smaller volume and consequently there is less numerical dissipation of magnetic energy. As a result, the jet spine gets more magnetised, $\sigma$ drops notably slower with radius, and the internal gas energy and pressure decrease with resolution.

%\subsubsection{Jet sheath}

We find that the jet-disk interface, the "sheath" is the most sensitive region to resolution. There are large fluctuations in the width of the jet sheath due to magnetic flux eruptions for $r\lesssim50r_g$, more prominent at higher resolutions. These fluctuations directly influence the variability of properties within the jet sheath, notably the temperature and magnetisation. At higher resolutions, we see more resolved wave-like features in the jet sheath, that may induce mixing between the sheath and plasma within $1 \lesssim \sigma \lesssim 3$ (see Fig.~\ref{fig:rho512} and \ref{fig:beta}). Mixing processes may include (1) shearing motion across the interface between the jet sheath and spine, (2) propagation of waves generated at the jet base, and (3) wobbling of the jet itself. Additionally for resolution $2240 \times 1056 \times 1024$ and $5376 \times 2304 \times 2304$, highly magnetised plasma from the jet spine supplies matter to the current sheets, and the reconnection exhaust deposits this hot plasma into the jet sheath \citep{Ripperda2022BlackReconnection}. At higher resolutions, we find that the jet sheath gets thinner, resulting in increased temperature, reduced magnetisation, and greater variability of $T_p$, $\hat{\sigma}$ and opening angle, while keeping an approximately constant total specific energy for all resolutions. These differences could affect the predicted multiwavelength spectra coming from the jet sheath, relevant to e.g. the EHT collaboration and VLBI imaging of large-scale jets. In the next paper on jet images, we will analyse whether a hotter, thinner jet sheath could result in increased limb-brightening in the observed images \citep[e.g.][]{Kim2018A&A...616A.188K,Janssen2021EventA} or potentially lead to filamentary structures observed in jets \citep{Fuentes2023NatAs...7.1359F}. %Formulating a function to describe how volume- and time-averaged jet sheath properties depend on resolution, similar to a "subgrid-scale model", is complex, highly setup-dependent, and beyond our paper's scope.
%The proton temperature in the jet sheath follows a less steep profile, $\propto  r^{-0.3}$, than in the disk $\propto  r^{-1.3}$. 
%For R1, the proton temperature remains nearly constant for radii between 100 and 500 $r_g$, and then follows a power law of $\propto r^{-0.3}$ for larger radii. This discrepancy arises from a lack of resolution in the polar direction at $r>100r_g$ along the jet axis for R1. %$T_{sheath}$ depends on the magnetisation in the jet spine, which is determined by numerical floors. $T_{sheath}$ is crucial to calculate synchrotron emission, as the sheath carries the highest density and has strong magnetic fields.

Capturing plasma mixing in flux tubes and along the jet-disk interface is essential for improved comparisons between observations and simulations. The presence of mixing instabilities could significantly influence the dissipation of energy, that could result in distinct multiwavelength emission from high-resolution simulations \citep[e.g.][]{Sironi2021ApJ...907L..44S,Zhdankin2023PhRvR...5d3023Z}. In forthcoming research we will investigate the influence of mixing instabilities on radiative emission in both total intensity and polarisation. We will focus on analysing non-thermal electron distribution functions on the dissipative regions. We will aim to simulate three-dimensional systems capable of accurately capturing jets extending up to $10^5r_g$ or more. This research is relevant to studies of AGN, such as \M87. By doing so, we aspire to deepen our comprehension of the fundamental horizon-scale physical processes and their implications for the dynamics and emission properties of larger-scale jets.

%% file: main.bbl
\begin{thebibliography}{}
\makeatletter
\relax
\def\mn@urlcharsother{\let\do\@makeother \do\$\do\&\do\#\do\^\do\_\do\%\do\~}
\def\mn@doi{\begingroup\mn@urlcharsother \@ifnextchar [ {\mn@doi@}
  {\mn@doi@[]}}
\def\mn@doi@[#1]#2{\def\@tempa{#1}\ifx\@tempa\@empty \href
  {http://dx.doi.org/#2} {doi:#2}\else \href {http://dx.doi.org/#2} {#1}\fi
  \endgroup}
\def\mn@eprint#1#2{\mn@eprint@#1:#2::\@nil}
\def\mn@eprint@arXiv#1{\href {http://arxiv.org/abs/#1} {{\tt arXiv:#1}}}
\def\mn@eprint@dblp#1{\href {http://dblp.uni-trier.de/rec/bibtex/#1.xml}
  {dblp:#1}}
\def\mn@eprint@#1:#2:#3:#4\@nil{\def\@tempa {#1}\def\@tempb {#2}\def\@tempc
  {#3}\ifx \@tempc \@empty \let \@tempc \@tempb \let \@tempb \@tempa \fi \ifx
  \@tempb \@empty \def\@tempb {arXiv}\fi \@ifundefined
  {mn@eprint@\@tempb}{\@tempb:\@tempc}{\expandafter \expandafter \csname
  mn@eprint@\@tempb\endcsname \expandafter{\@tempc}}}

\bibitem[\protect\citeauthoryear{{Acciari} et~al.,}{{Acciari}
  et~al.}{2010}]{Acciari2010ApJ...716..819A}
{Acciari} V.~A.,  et~al., 2010, \mn@doi [\apj] {10.1088/0004-637X/716/1/819},
  \href {https://ui.adsabs.harvard.edu/abs/2010ApJ...716..819A} {716, 819}

\bibitem[\protect\citeauthoryear{{Aharonian} et~al.,}{{Aharonian}
  et~al.}{2006}]{Aharonian2006Sci...314.1424A}
{Aharonian} F.,  et~al., 2006, \mn@doi [Science] {10.1126/science.1134408},
  \href {https://ui.adsabs.harvard.edu/abs/2006Sci...314.1424A} {314, 1424}

\bibitem[\protect\citeauthoryear{{Akiyama} et~al.,}{{Akiyama}
  et~al.}{2023}]{EHTC2023IX}
{Akiyama} K.,  et~al., 2023, \mn@doi [\apjl] {10.3847/2041-8213/acff70}, \href
  {https://ui.adsabs.harvard.edu/abs/2023ApJ...957L..20A} {957, L20}

\bibitem[\protect\citeauthoryear{{Aliu} et~al.,}{{Aliu}
  et~al.}{2012}]{Aliu2012ApJ...746..141A}
{Aliu} E.,  et~al., 2012, \mn@doi [\apj] {10.1088/0004-637X/746/2/141}, \href
  {https://ui.adsabs.harvard.edu/abs/2012ApJ...746..141A} {746, 141}

\bibitem[\protect\citeauthoryear{{Anninos}, {Fragile}  \&
  {Salmonson}}{{Anninos} et~al.}{2005}]{Anninos_2005_COSMOS++}
{Anninos} P.,  {Fragile} P.~C.,   {Salmonson} J.~D.,  2005, \mn@doi [\apj]
  {10.1086/497294}, \href
  {https://ui.adsabs.harvard.edu/abs/2005ApJ...635..723A} {635, 723}

\bibitem[\protect\citeauthoryear{{Asada} \& {Nakamura}}{{Asada} \&
  {Nakamura}}{2012}]{AsadaNakamura2012ApJ...745L..28A}
{Asada} K.,  {Nakamura} M.,  2012, \mn@doi [\apjl]
  {10.1088/2041-8205/745/2/L28}, \href
  {https://ui.adsabs.harvard.edu/abs/2012ApJ...745L..28A} {745, L28}

\bibitem[\protect\citeauthoryear{{Baganoff} et~al.,}{{Baganoff}
  et~al.}{2001}]{Baganoff2001Natur.413...45B}
{Baganoff} F.~K.,  et~al., 2001, \mn@doi [\nat] {10.1038/35092510}, \href
  {https://ui.adsabs.harvard.edu/abs/2001Natur.413...45B} {413, 45}

\bibitem[\protect\citeauthoryear{Balbus \& Hawley}{Balbus \&
  Hawley}{1991}]{MRIBalbus}
Balbus S.~A.,  Hawley J.~F.,  1991, \mn@doi [Astrophysical Journal]
  {10.1086/170270}, 376, 214

\bibitem[\protect\citeauthoryear{{Balbus} \& {Hawley}}{{Balbus} \&
  {Hawley}}{1998}]{Balbus_Hawley_1998}
{Balbus} S.~A.,  {Hawley} J.~F.,  1998, \mn@doi [Reviews of Modern Physics]
  {10.1103/RevModPhys.70.1}, \href
  {https://ui.adsabs.harvard.edu/abs/1998RvMP...70....1B} {70, 1}

\bibitem[\protect\citeauthoryear{{Begelman}, {Scepi}  \& {Dexter}}{{Begelman}
  et~al.}{2022}]{Begelman2022MNRAS.511.2040B}
{Begelman} M.~C.,  {Scepi} N.,   {Dexter} J.,  2022, \mn@doi [\mnras]
  {10.1093/mnras/stab3790}, \href
  {https://ui.adsabs.harvard.edu/abs/2022MNRAS.511.2040B} {511, 2040}

\bibitem[\protect\citeauthoryear{Bhattacharjee, Huang, Yang  \&
  Rogers}{Bhattacharjee et~al.}{2009}]{Bhattacharjee2009FastInstabilities}
Bhattacharjee A.,  Huang Y.-M.,  Yang H.,   Rogers B.,  2009, \mn@doi [Physics
  of Plasmas] {10.1063/1.3264103}, 16, 112102

\bibitem[\protect\citeauthoryear{{Bisnovatyi-Kogan} \&
  {Ruzmaikin}}{{Bisnovatyi-Kogan} \&
  {Ruzmaikin}}{1974}]{Bisnovatyi-Kogan1974Ap&SS..28...45B}
{Bisnovatyi-Kogan} G.~S.,  {Ruzmaikin} A.~A.,  1974, \mn@doi [\apss]
  {10.1007/BF00642237}, \href
  {https://ui.adsabs.harvard.edu/abs/1974Ap&SS..28...45B} {28, 45}

\bibitem[\protect\citeauthoryear{Blanch}{Blanch}{2021}]{blanch2021magic}
Blanch O.,  2021, The Astronomer's Telegram, 14483, 1

\bibitem[\protect\citeauthoryear{Blandford \& Znajek}{Blandford \&
  Znajek}{1977}]{blandford1977electromagnetic}
Blandford R.~D.,  Znajek R.~L.,  1977, Monthly Notices of the Royal
  Astronomical Society, 179, 433

\bibitem[\protect\citeauthoryear{{Bransgrove}, {Ripperda}  \&
  {Philippov}}{{Bransgrove} et~al.}{2021}]{Bransgrove2021PhRvL.127e5101B}
{Bransgrove} A.,  {Ripperda} B.,   {Philippov} A.,  2021, \mn@doi [\prl]
  {10.1103/PhysRevLett.127.055101}, \href
  {https://ui.adsabs.harvard.edu/abs/2021PhRvL.127e5101B} {127, 055101}

\bibitem[\protect\citeauthoryear{{Chatterjee} \& {Narayan}}{{Chatterjee} \&
  {Narayan}}{2022}]{ChatterjeeNarayan2022ApJ...941...30C}
{Chatterjee} K.,  {Narayan} R.,  2022, \mn@doi [\apj]
  {10.3847/1538-4357/ac9d97}, \href
  {https://ui.adsabs.harvard.edu/abs/2022ApJ...941...30C} {941, 30}

\bibitem[\protect\citeauthoryear{Chatterjee, Liska, Tchekhovskoy  \&
  Markoff}{Chatterjee et~al.}{2019}]{JetsKoushik}
Chatterjee K.,  Liska M.,  Tchekhovskoy A.,   Markoff S.~B.,  2019, \mn@doi
  [Monthly Notices of the Royal Astronomical Society] {10.1093/mnras/stz2626},
  490, 2200

\bibitem[\protect\citeauthoryear{Colella \& Woodward}{Colella \&
  Woodward}{1984}]{Colella1984}
Colella P.,  Woodward P.~R.,  1984, \mn@doi [JCoPh]
  {10.1016/0021-9991(84)90143-8}, 54, 174

\bibitem[\protect\citeauthoryear{Davelaar, Mo{\'s}cibrodzka, Bronzwaer  \&
  Falcke}{Davelaar et~al.}{2018}]{davelaar2018general}
Davelaar J.,  Mo{\'s}cibrodzka M.,  Bronzwaer T.,   Falcke H.,  2018, Astronomy
  \& Astrophysics, 612, A34

\bibitem[\protect\citeauthoryear{{Davelaar} et~al.,}{{Davelaar}
  et~al.}{2023}]{Davelaar2023ApJ...959L...3D}
{Davelaar} J.,  et~al., 2023, \mn@doi [\apjl] {10.3847/2041-8213/ad0b79}, \href
  {https://ui.adsabs.harvard.edu/abs/2023ApJ...959L...3D} {959, L3}

\bibitem[\protect\citeauthoryear{{Del Zanna}, {Zanotti}, {Bucciantini}  \&
  {Londrillo}}{{Del Zanna} et~al.}{2007}]{Del_Zanna_2007_ECHO}
{Del Zanna} L.,  {Zanotti} O.,  {Bucciantini} N.,   {Londrillo} P.,  2007,
  \mn@doi [\aap] {10.1051/0004-6361:20077093}, \href
  {https://ui.adsabs.harvard.edu/abs/2007A&A...473...11D} {473, 11}

\bibitem[\protect\citeauthoryear{{Dexter} et~al.,}{{Dexter}
  et~al.}{2020}]{DexterTchekhovskoy2020MNRAS.497.4999D}
{Dexter} J.,  et~al., 2020, \mn@doi [\mnras] {10.1093/mnras/staa2288}, \href
  {https://ui.adsabs.harvard.edu/abs/2020MNRAS.497.4999D} {497, 4999}

\bibitem[\protect\citeauthoryear{{EHT MWL Science Working Group}, {Algaba},
  {Anczarski}, {Asada}, {Balokovi{\'c}}, {Chandra}  \& et al.}{{EHT MWL Science
  Working Group} et~al.}{2021}]{EHTMWL2021}
{EHT MWL Science Working Group} {Algaba} J.~C.,  {Anczarski} J.,  {Asada} K.,
  {Balokovi{\'c}} M.,  {Chandra} S.,   et al. 2021, \mn@doi [\apjl]
  {10.3847/2041-8213/abef71}, \href
  {https://ui.adsabs.harvard.edu/abs/2021ApJ...911L..11E} {911, L11}

\bibitem[\protect\citeauthoryear{{Eckart} et~al.,}{{Eckart}
  et~al.}{2004}]{Eckart2004A&A...427....1E}
{Eckart} A.,  et~al., 2004, \mn@doi [\aap] {10.1051/0004-6361:20040495}, \href
  {https://ui.adsabs.harvard.edu/abs/2004A&A...427....1E} {427, 1}

\bibitem[\protect\citeauthoryear{{Etienne}, {Paschalidis}, {Haas}, {M{\"o}sta}
  \& {Shapiro}}{{Etienne} et~al.}{2015}]{Etienne_2015_IllinoisGRMHD}
{Etienne} Z.~B.,  {Paschalidis} V.,  {Haas} R.,  {M{\"o}sta} P.,   {Shapiro}
  S.~L.,  2015, \mn@doi [Classical and Quantum Gravity]
  {10.1088/0264-9381/32/17/175009}, \href
  {https://ui.adsabs.harvard.edu/abs/2015CQGra..32q5009E} {32, 175009}

\bibitem[\protect\citeauthoryear{{Event Horizon Telescope Collaboration},
  {Akiyama}, {Alberdi}, {Alef}, {Asada}  \& et al.}{{Event Horizon Telescope
  Collaboration} et~al.}{2019a}]{EHTC2019I}
{Event Horizon Telescope Collaboration} {Akiyama} K.,  {Alberdi} A.,  {Alef}
  W.,  {Asada} K.,   et al. 2019a, \mn@doi [\apjl] {10.3847/2041-8213/ab0ec7},
  \href {https://ui.adsabs.harvard.edu/abs/2019ApJ...875L...1E} {875, L1}

\bibitem[\protect\citeauthoryear{{Event Horizon Telescope Collaboration},
  {Akiyama}, {Alberdi}, {Alef}, {Asada}  \& et al.}{{Event Horizon Telescope
  Collaboration} et~al.}{2019b}]{EHTC2019V}
{Event Horizon Telescope Collaboration} {Akiyama} K.,  {Alberdi} A.,  {Alef}
  W.,  {Asada} K.,   et al. 2019b, \mn@doi [\apjl] {10.3847/2041-8213/ab0f43},
  \href {https://ui.adsabs.harvard.edu/abs/2019ApJ...875L...5E} {875, L5}

\bibitem[\protect\citeauthoryear{{Event Horizon Telescope Collaboration},
  {Akiyama}, {Algaba}, {Alberdi}, {Alef}  \& et al.}{{Event Horizon Telescope
  Collaboration} et~al.}{2021}]{EHTC2021VIII}
{Event Horizon Telescope Collaboration} {Akiyama} K.,  {Algaba} J.~C.,
  {Alberdi} A.,  {Alef} W.,   et al. 2021, \mn@doi [\apjl]
  {10.3847/2041-8213/abe4de}, \href
  {https://ui.adsabs.harvard.edu/abs/2021ApJ...910L..13E} {910, L13}

\bibitem[\protect\citeauthoryear{{Event Horizon Telescope Collaboration},
  {Akiyama}, {Alberdi}, {Alef}, {Algaba}  \& et al.}{{Event Horizon Telescope
  Collaboration} et~al.}{2022a}]{EHTC2022I}
{Event Horizon Telescope Collaboration} {Akiyama} K.,  {Alberdi} A.,  {Alef}
  W.,  {Algaba} J.~C.,   et al. 2022a, \mn@doi [\apjl]
  {10.3847/2041-8213/ac6674}, \href
  {https://ui.adsabs.harvard.edu/abs/2022ApJ...930L..12E} {930, L12}

\bibitem[\protect\citeauthoryear{{Event Horizon Telescope Collaboration},
  {Akiyama}, {Alberdi}, {Alef}, {Algaba}  \& et al.}{{Event Horizon Telescope
  Collaboration} et~al.}{2022b}]{EHTC2022II}
{Event Horizon Telescope Collaboration} {Akiyama} K.,  {Alberdi} A.,  {Alef}
  W.,  {Algaba} J.~C.,   et al. 2022b, \mn@doi [\apjl]
  {10.3847/2041-8213/ac6675}, \href
  {https://ui.adsabs.harvard.edu/abs/2022ApJ...930L..13E} {930, L13}

\bibitem[\protect\citeauthoryear{{Event Horizon Telescope Collaboration},
  {Akiyama}, {Alberdi}, {Alef}, {Algaba}  \& et al.}{{Event Horizon Telescope
  Collaboration} et~al.}{2022c}]{EHTC2022V}
{Event Horizon Telescope Collaboration} {Akiyama} K.,  {Alberdi} A.,  {Alef}
  W.,  {Algaba} J.~C.,   et al. 2022c, \mn@doi [\apjl]
  {10.3847/2041-8213/ac6672}, \href
  {https://ui.adsabs.harvard.edu/abs/2022ApJ...930L..16E} {930, L16}

\bibitem[\protect\citeauthoryear{Fabian}{Fabian}{2012}]{fabian2012observational}
Fabian A.,  2012, Annual Review of Astronomy and Astrophysics, 50, 455

\bibitem[\protect\citeauthoryear{Fishbone \& Moncrief}{Fishbone \&
  Moncrief}{1976}]{Fishbone1976RelativisticHoles}
Fishbone L.~G.,  Moncrief V.,  1976, The Astrophysical Journal, 207, 962

\bibitem[\protect\citeauthoryear{{Fuentes} et~al.,}{{Fuentes}
  et~al.}{2023}]{Fuentes2023NatAs...7.1359F}
{Fuentes} A.,  et~al., 2023, \mn@doi [Nature Astronomy]
  {10.1038/s41550-023-02105-7}, \href
  {https://ui.adsabs.harvard.edu/abs/2023NatAs...7.1359F} {7, 1359}

\bibitem[\protect\citeauthoryear{{GRAVITY Collaboration} et~al.,}{{GRAVITY
  Collaboration} et~al.}{2018}]{GRAVITY2018A&A...618L..10G}
{GRAVITY Collaboration} et~al., 2018, \mn@doi [\aap]
  {10.1051/0004-6361/201834294}, \href
  {https://ui.adsabs.harvard.edu/abs/2018A&A...618L..10G} {618, L10}

\bibitem[\protect\citeauthoryear{Gammie}{Gammie}{2004}]{Gammie2004TheMetric}
Gammie C.~F.,  2004, The Astrophysical Journal, 614

\bibitem[\protect\citeauthoryear{{Gammie}, {McKinney}  \& {T{\'o}th}}{{Gammie}
  et~al.}{2003}]{Gammie2003_HARM}
{Gammie} C.~F.,  {McKinney} J.~C.,   {T{\'o}th} G.,  2003, \mn@doi [\apj]
  {10.1086/374594}, \href
  {https://ui.adsabs.harvard.edu/abs/2003ApJ...589..444G} {589, 444}

\bibitem[\protect\citeauthoryear{{Gardiner} \& {Stone}}{{Gardiner} \&
  {Stone}}{2005}]{Gardiner2005JCoPh}
{Gardiner} T.~A.,  {Stone} J.~M.,  2005, \mn@doi [Journal of Computational
  Physics] {10.1016/j.jcp.2004.11.016}, \href
  {https://ui.adsabs.harvard.edu/abs/2005JCoPh.205..509G} {205, 509}

\bibitem[\protect\citeauthoryear{{Goedbloed} \& {Keppens}}{{Goedbloed} \&
  {Keppens}}{2022}]{Goedbloed2022ApJS..259...65G}
{Goedbloed} H.,  {Keppens} R.,  2022, \mn@doi [\apjs]
  {10.3847/1538-4365/ac573c}, \href
  {https://ui.adsabs.harvard.edu/abs/2022ApJS..259...65G} {259, 65}

\bibitem[\protect\citeauthoryear{Gravity~Collaboration Abuter
  et~al.,}{Gravity~Collaboration et~al.}{2021}]{abuter2021constraining}
Gravity~Collaboration Abuter R.,  et~al., 2021, Astronomy and
  Astrophysics-A\&A, 654, A22

\bibitem[\protect\citeauthoryear{{Grete}, {Glines}  \& {O'Shea}}{{Grete}
  et~al.}{2021}]{Grete2021ITPDS..32...85G}
{Grete} P.,  {Glines} F.~W.,   {O'Shea} B.~W.,  2021, \mn@doi [IEEE
  Transactions on Parallel and Distributed Systems]
  {10.1109/TPDS.2020.3010016}, \href
  {https://ui.adsabs.harvard.edu/abs/2021ITPDS..32...85G} {32, 85}

\bibitem[\protect\citeauthoryear{{Guan} \& {Gammie}}{{Guan} \&
  {Gammie}}{2008}]{Guan2008ApJS..174..145G}
{Guan} X.,  {Gammie} C.~F.,  2008, \mn@doi [\apjs] {10.1086/521147}, \href
  {https://ui.adsabs.harvard.edu/abs/2008ApJS..174..145G} {174, 145}

\bibitem[\protect\citeauthoryear{{Hada} et~al.,}{{Hada}
  et~al.}{2013}]{Hada2013ApJ...775...70H}
{Hada} K.,  et~al., 2013, \mn@doi [\apj] {10.1088/0004-637X/775/1/70}, \href
  {https://ui.adsabs.harvard.edu/abs/2013ApJ...775...70H} {775, 70}

\bibitem[\protect\citeauthoryear{{Hakobyan}, {Ripperda}  \&
  {Philippov}}{{Hakobyan} et~al.}{2023}]{Hakobyan2023ApJ...943L..29H}
{Hakobyan} H.,  {Ripperda} B.,   {Philippov} A.~A.,  2023, \mn@doi [\apjl]
  {10.3847/2041-8213/acb264}, \href
  {https://ui.adsabs.harvard.edu/abs/2023ApJ...943L..29H} {943, L29}

\bibitem[\protect\citeauthoryear{Hide \& Palmer}{Hide \&
  Palmer}{1982}]{hide1982generalization}
Hide R.,  Palmer T.,  1982, Geophysical \& Astrophysical Fluid Dynamics, 19,
  301

\bibitem[\protect\citeauthoryear{{Igumenshchev}}{{Igumenshchev}}{2008}]{Igumenshchev2008ApJ...677..317I}
{Igumenshchev} I.~V.,  2008, \mn@doi [\apj] {10.1086/529025}, \href
  {https://ui.adsabs.harvard.edu/abs/2008ApJ...677..317I} {677, 317}

\bibitem[\protect\citeauthoryear{{Igumenshchev}, {Narayan}  \&
  {Abramowicz}}{{Igumenshchev} et~al.}{2003}]{Igumenshchev2003ApJ...592.1042I}
{Igumenshchev} I.~V.,  {Narayan} R.,   {Abramowicz} M.~A.,  2003, \mn@doi
  [\apj] {10.1086/375769}, \href
  {https://ui.adsabs.harvard.edu/abs/2003ApJ...592.1042I} {592, 1042}

\bibitem[\protect\citeauthoryear{Janssen et~al.,}{Janssen
  et~al.}{2021}]{Janssen2021EventA}
Janssen M.,  et~al., 2021, \mn@doi [Nature Astronomy]
  {10.1038/s41550-021-01417-w}, 5, 1017

\bibitem[\protect\citeauthoryear{{Jia}, {Ripperda}, {Quataert}, {White},
  {Chatterjee}, {Philippov}  \& {Liska}}{{Jia}
  et~al.}{2023}]{Jia2023MNRAS.526.2924J}
{Jia} H.,  {Ripperda} B.,  {Quataert} E.,  {White} C.~J.,  {Chatterjee} K.,
  {Philippov} A.,   {Liska} M.,  2023, \mn@doi [\mnras]
  {10.1093/mnras/stad2935}, \href
  {https://ui.adsabs.harvard.edu/abs/2023MNRAS.526.2924J} {526, 2924}

\bibitem[\protect\citeauthoryear{{Kim} et~al.,}{{Kim}
  et~al.}{2018}]{Kim2018A&A...616A.188K}
{Kim} J.~Y.,  et~al., 2018, \mn@doi [\aap] {10.1051/0004-6361/201832921}, \href
  {https://ui.adsabs.harvard.edu/abs/2018A&A...616A.188K} {616, A188}

\bibitem[\protect\citeauthoryear{{Liska}, {Tchekhovskoy}, {Ingram}  \& {van der
  Klis}}{{Liska}
  et~al.}{2019}]{Liska_Tchekhovskoy_Ingram2019_Bardeen-Petterson}
{Liska} M.,  {Tchekhovskoy} A.,  {Ingram} A.,   {van der Klis} M.,  2019,
  \mn@doi [\mnras] {10.1093/mnras/stz834}, \href
  {https://ui.adsabs.harvard.edu/abs/2019MNRAS.487..550L} {487, 550}

\bibitem[\protect\citeauthoryear{{Liska}, {Tchekhovskoy}  \&
  {Quataert}}{{Liska} et~al.}{2020}]{Liska2020MNRAS.494.3656L}
{Liska} M.,  {Tchekhovskoy} A.,   {Quataert} E.,  2020, \mn@doi [\mnras]
  {10.1093/mnras/staa955}, \href
  {https://ui.adsabs.harvard.edu/abs/2020MNRAS.494.3656L} {494, 3656}

\bibitem[\protect\citeauthoryear{{Liska} et~al.,}{{Liska}
  et~al.}{2022}]{Liska2022ApJS..263...26L}
{Liska} M.~T.~P.,  et~al., 2022, \mn@doi [\apjs] {10.3847/1538-4365/ac9966},
  \href {https://ui.adsabs.harvard.edu/abs/2022ApJS..263...26L} {263, 26}

\bibitem[\protect\citeauthoryear{{Liska}, {Kaaz}, {Chatterjee}, {Emami}  \&
  {Musoke}}{{Liska} et~al.}{2024}]{Liska2024ApJ...966...47L}
{Liska} M.~T.~P.,  {Kaaz} N.,  {Chatterjee} K.,  {Emami} R.,   {Musoke} G.,
  2024, \mn@doi [\apj] {10.3847/1538-4357/ad344a}, \href
  {https://ui.adsabs.harvard.edu/abs/2024ApJ...966...47L} {966, 47}

\bibitem[\protect\citeauthoryear{{McKinney}, {Tchekhovskoy}  \&
  {Blandford}}{{McKinney} et~al.}{2012}]{McKinney2012MNRAS.423.3083M}
{McKinney} J.~C.,  {Tchekhovskoy} A.,   {Blandford} R.~D.,  2012, \mn@doi
  [\mnras] {10.1111/j.1365-2966.2012.21074.x}, \href
  {https://ui.adsabs.harvard.edu/abs/2012MNRAS.423.3083M} {423, 3083}

\bibitem[\protect\citeauthoryear{{Mertens}, {Lobanov}, {Walker}  \&
  {Hardee}}{{Mertens} et~al.}{2016}]{Mertens2016A&A...595A..54M}
{Mertens} F.,  {Lobanov} A.~P.,  {Walker} R.~C.,   {Hardee} P.~E.,  2016,
  \mn@doi [\aap] {10.1051/0004-6361/201628829}, \href
  {https://ui.adsabs.harvard.edu/abs/2016A&A...595A..54M} {595, A54}

\bibitem[\protect\citeauthoryear{Mo{\'s}cibrodzka \& Falcke}{Mo{\'s}cibrodzka
  \& Falcke}{2013}]{moscibrodzka2013coupled}
Mo{\'s}cibrodzka M.,  Falcke H.,  2013, Astronomy \& Astrophysics, 559, L3

\bibitem[\protect\citeauthoryear{{Najafi-Ziyazi}, {Davelaar}, {Mizuno}  \&
  {Porth}}{{Najafi-Ziyazi} et~al.}{2024}]{Najafi-Ziyazi2024MNRAS.531.3961N}
{Najafi-Ziyazi} M.,  {Davelaar} J.,  {Mizuno} Y.,   {Porth} O.,  2024, \mn@doi
  [\mnras] {10.1093/mnras/stae1343}, \href
  {https://ui.adsabs.harvard.edu/abs/2024MNRAS.531.3961N} {531, 3961}

\bibitem[\protect\citeauthoryear{{Nakamura} et~al.,}{{Nakamura}
  et~al.}{2018}]{Nakamura2018ApJ...868..146N}
{Nakamura} M.,  et~al., 2018, \mn@doi [\apj] {10.3847/1538-4357/aaeb2d}, \href
  {https://ui.adsabs.harvard.edu/abs/2018ApJ...868..146N} {868, 146}

\bibitem[\protect\citeauthoryear{{Narayan}, {Igumenshchev}  \&
  {Abramowicz}}{{Narayan} et~al.}{2003}]{Narayan2003PASJ...55L..69N}
{Narayan} R.,  {Igumenshchev} I.~V.,   {Abramowicz} M.~A.,  2003, \mn@doi
  [\pasj] {10.1093/pasj/55.6.L69}, \href
  {https://ui.adsabs.harvard.edu/abs/2003PASJ...55L..69N} {55, L69}

\bibitem[\protect\citeauthoryear{Narayan, Sadowski, Penna  \& Kulkarni}{Narayan
  et~al.}{2012}]{Narayan2012GRMHDOutflows}
Narayan R.,  Sadowski A.,  Penna R.~F.,   Kulkarni A.~K.,  2012, \mn@doi
  [Monthly Notices of the Royal Astronomical Society]
  {10.1111/j.1365-2966.2012.22002.x}, 426, 3241

\bibitem[\protect\citeauthoryear{{Narayan}, {Chael}, {Chatterjee}, {Ricarte}
  \& {Curd}}{{Narayan} et~al.}{2022}]{Narayan2022MNRAS.511.3795N}
{Narayan} R.,  {Chael} A.,  {Chatterjee} K.,  {Ricarte} A.,   {Curd} B.,  2022,
  \mn@doi [\mnras] {10.1093/mnras/stac285}, \href
  {https://ui.adsabs.harvard.edu/abs/2022MNRAS.511.3795N} {511, 3795}

\bibitem[\protect\citeauthoryear{Neilsen et~al.,}{Neilsen
  et~al.}{2015}]{neilsen2015x}
Neilsen J.,  et~al., 2015, The Astrophysical Journal, 799, 199

\bibitem[\protect\citeauthoryear{Noble, Gammie, McKinney  \& Zanna}{Noble
  et~al.}{2006}]{Noble2006}
Noble S.~C.,  Gammie C.~F.,  McKinney J.~C.,   Zanna L.~D.,  2006, \mn@doi [The
  Astrophysical Journal] {10.1086/500349}, 641, 626

\bibitem[\protect\citeauthoryear{{Pessah}, {Chan}  \& {Psaltis}}{{Pessah}
  et~al.}{2006}]{Pessah2006MNRAS.372..183P}
{Pessah} M.~E.,  {Chan} C.-K.,   {Psaltis} D.,  2006, \mn@doi [\mnras]
  {10.1111/j.1365-2966.2006.10824.x}, \href
  {https://ui.adsabs.harvard.edu/abs/2006MNRAS.372..183P} {372, 183}

\bibitem[\protect\citeauthoryear{{Porth} \& {Komissarov}}{{Porth} \&
  {Komissarov}}{2015}]{Porth2015MNRAS.452.1089P}
{Porth} O.,  {Komissarov} S.~S.,  2015, \mn@doi [\mnras]
  {10.1093/mnras/stv1295}, \href
  {https://ui.adsabs.harvard.edu/abs/2015MNRAS.452.1089P} {452, 1089}

\bibitem[\protect\citeauthoryear{{Porth}, {Olivares}, {Mizuno}, {Younsi},
  {Rezzolla}, {Moscibrodzka}, {Falcke}  \& {Kramer}}{{Porth}
  et~al.}{2017}]{Porth_2017_BHAC}
{Porth} O.,  {Olivares} H.,  {Mizuno} Y.,  {Younsi} Z.,  {Rezzolla} L.,
  {Moscibrodzka} M.,  {Falcke} H.,   {Kramer} M.,  2017, \mn@doi [Computational
  Astrophysics and Cosmology] {10.1186/s40668-017-0020-2}, \href
  {https://ui.adsabs.harvard.edu/abs/2017ComAC...4....1P} {4, 1}

\bibitem[\protect\citeauthoryear{Porth et~al.,}{Porth
  et~al.}{2019}]{EHTGRMHDcomparison}
Porth O.,  et~al., 2019, \mn@doi [The Astrophysical Journal Supplement Series]
  {10.3847/1538-4365/ab29fd}, 243, 26

\bibitem[\protect\citeauthoryear{{Porth}, {Mizuno}, {Younsi}  \&
  {Fromm}}{{Porth} et~al.}{2021}]{Porth2021MNRAS.502.2023P}
{Porth} O.,  {Mizuno} Y.,  {Younsi} Z.,   {Fromm} C.~M.,  2021, \mn@doi
  [\mnras] {10.1093/mnras/stab163}, \href
  {https://ui.adsabs.harvard.edu/abs/2021MNRAS.502.2023P} {502, 2023}

\bibitem[\protect\citeauthoryear{{Prather}, {Wong}, {Dhruv}, {Ryan}, {Dolence},
  {Ressler}  \& {Gammie}}{{Prather} et~al.}{2021}]{Prather2021JOSS....6.3336P}
{Prather} B.,  {Wong} G.,  {Dhruv} V.,  {Ryan} B.,  {Dolence} J.,  {Ressler}
  S.,   {Gammie} C.,  2021, \mn@doi [The Journal of Open Source Software]
  {10.21105/joss.03336}, \href
  {https://ui.adsabs.harvard.edu/abs/2021JOSS....6.3336P} {6, 3336}

\bibitem[\protect\citeauthoryear{{Ressler}, {White}  \& {Quataert}}{{Ressler}
  et~al.}{2023}]{Ressler2023MNRAS.521.4277R}
{Ressler} S.~M.,  {White} C.~J.,   {Quataert} E.,  2023, \mn@doi [\mnras]
  {10.1093/mnras/stad837}, \href
  {https://ui.adsabs.harvard.edu/abs/2023MNRAS.521.4277R} {521, 4277}

\bibitem[\protect\citeauthoryear{{Ripperda} et~al.,}{{Ripperda}
  et~al.}{2019}]{Ripperda2019ApJS..244...10R}
{Ripperda} B.,  et~al., 2019, \mn@doi [\apjs] {10.3847/1538-4365/ab3922}, \href
  {https://ui.adsabs.harvard.edu/abs/2019ApJS..244...10R} {244, 10}

\bibitem[\protect\citeauthoryear{{Ripperda}, {Bacchini}  \&
  {Philippov}}{{Ripperda} et~al.}{2020}]{Ripperda2020ApJ...900..100R}
{Ripperda} B.,  {Bacchini} F.,   {Philippov} A.~A.,  2020, \mn@doi [\apj]
  {10.3847/1538-4357/ababab}, \href
  {https://ui.adsabs.harvard.edu/abs/2020ApJ...900..100R} {900, 100}

\bibitem[\protect\citeauthoryear{{Ripperda}, {Liska}, {Chatterjee}, {Musoke},
  {Philippov}, {Markoff}, {Tchekhovskoy}  \& {Younsi}}{{Ripperda}
  et~al.}{2022}]{Ripperda2022BlackReconnection}
{Ripperda} B.,  {Liska} M.,  {Chatterjee} K.,  {Musoke} G.,  {Philippov} A.~A.,
   {Markoff} S.~B.,  {Tchekhovskoy} A.,   {Younsi} Z.,  2022, \mn@doi [\apjl]
  {10.3847/2041-8213/ac46a1}, \href
  {https://ui.adsabs.harvard.edu/abs/2022ApJ...924L..32R} {924, L32}

\bibitem[\protect\citeauthoryear{Schawinski, Thomas, Sarzi, Maraston, Kaviraj,
  Joo, Yi  \& Silk}{Schawinski et~al.}{2007}]{schawinski2007observational}
Schawinski K.,  Thomas D.,  Sarzi M.,  Maraston C.,  Kaviraj S.,  Joo S.-J.,
  Yi S.~K.,   Silk J.,  2007, Monthly Notices of the Royal Astronomical
  Society, 382, 1415

\bibitem[\protect\citeauthoryear{Shakura \& Sunyaev}{Shakura \&
  Sunyaev}{1973}]{shakura1973black}
Shakura N.~I.,  Sunyaev R.~A.,  1973, Astronomy and Astrophysics, 24, 337

\bibitem[\protect\citeauthoryear{Silk \& Rees}{Silk \&
  Rees}{1998}]{silk1998quasars}
Silk J.,  Rees M.~J.,  1998, Astronomy and Astrophysics, 331, L1

\bibitem[\protect\citeauthoryear{{Sironi}, {Rowan}  \& {Narayan}}{{Sironi}
  et~al.}{2021}]{Sironi2021ApJ...907L..44S}
{Sironi} L.,  {Rowan} M.~E.,   {Narayan} R.,  2021, \mn@doi [\apjl]
  {10.3847/2041-8213/abd9bc}, \href
  {https://ui.adsabs.harvard.edu/abs/2021ApJ...907L..44S} {907, L44}

\bibitem[\protect\citeauthoryear{{S{\k{a}}dowski}, {Narayan}, {McKinney}  \&
  {Tchekhovskoy}}{{S{\k{a}}dowski} et~al.}{2014}]{Sadowski2014_KORAL}
{S{\k{a}}dowski} A.,  {Narayan} R.,  {McKinney} J.~C.,   {Tchekhovskoy} A.,
  2014, \mn@doi [\mnras] {10.1093/mnras/stt2479}, \href
  {https://ui.adsabs.harvard.edu/abs/2014MNRAS.439..503S} {439, 503}

\bibitem[\protect\citeauthoryear{Stanzione, West, Evans, Minyard, Ghattas  \&
  Panda}{Stanzione et~al.}{2020}]{Stanzione2020Frontera:Foundation}
Stanzione D.,  West J.,  Evans R.~T.,  Minyard T.,  Ghattas O.,   Panda D.~K.,
  2020, in Practice and Experience in Advanced Research Computing. ACM, New
  York, NY, USA, pp 106--111, \mn@doi{10.1145/3311790.3396656}

\bibitem[\protect\citeauthoryear{{Takahashi}, {Ohsuga}, {Kawashima}  \&
  {Sekiguchi}}{{Takahashi} et~al.}{2016}]{Takahashi_2016ApJ_Uwabami}
{Takahashi} H.~R.,  {Ohsuga} K.,  {Kawashima} T.,   {Sekiguchi} Y.,  2016,
  \mn@doi [\apj] {10.3847/0004-637X/826/1/23}, \href
  {https://ui.adsabs.harvard.edu/abs/2016ApJ...826...23T} {826, 23}

\bibitem[\protect\citeauthoryear{Tchekhovskoy, Narayan  \&
  Mckinney}{Tchekhovskoy et~al.}{2011}]{Tchekhovskoy2011EfficientHole}
Tchekhovskoy A.,  Narayan R.,   Mckinney J.~C.,  2011, \mn@doi [Monthly Notices
  of the Royal Astronomical Society: Letters]
  {10.1111/j.1745-3933.2011.01147.x}, 418

\bibitem[\protect\citeauthoryear{Tchekhovskoy, McKinney  \&
  Narayan}{Tchekhovskoy et~al.}{2012}]{tchekhovskoy2012general}
Tchekhovskoy A.,  McKinney J.~C.,   Narayan R.,  2012, in Journal of Physics:
  Conference Series. p. 012040

\bibitem[\protect\citeauthoryear{{Uzdensky}, {Loureiro}  \&
  {Schekochihin}}{{Uzdensky} et~al.}{2010}]{Uzdensky2010PhRvL.105w5002U}
{Uzdensky} D.~A.,  {Loureiro} N.~F.,   {Schekochihin} A.~A.,  2010, \mn@doi
  [\prl] {10.1103/PhysRevLett.105.235002}, \href
  {https://ui.adsabs.harvard.edu/abs/2010PhRvL.105w5002U} {105, 235002}

\bibitem[\protect\citeauthoryear{White, Stone  \& Quataert}{White
  et~al.}{2019}]{MADresWhite}
White C.~J.,  Stone J.~M.,   Quataert E.,  2019, \mn@doi [The Astrophysical
  Journal] {10.3847/1538-4357/ab0c0c}, 874, 168

\bibitem[\protect\citeauthoryear{{White}, {Mullen}, {Jiang}, {Davis}, {Stone},
  {Morozova}  \& {Zhang}}{{White}
  et~al.}{2023}]{WhiteAthena++2023ApJ...949..103W}
{White} C.~J.,  {Mullen} P.~D.,  {Jiang} Y.-F.,  {Davis} S.~W.,  {Stone} J.~M.,
   {Morozova} V.,   {Zhang} L.,  2023, \mn@doi [\apj]
  {10.3847/1538-4357/acc8cf}, \href
  {https://ui.adsabs.harvard.edu/abs/2023ApJ...949..103W} {949, 103}

\bibitem[\protect\citeauthoryear{{Wong}, {Du}, {Prather}  \& {Gammie}}{{Wong}
  et~al.}{2021}]{Wong2021ApJ...914...55W}
{Wong} G.~N.,  {Du} Y.,  {Prather} B.~S.,   {Gammie} C.~F.,  2021, \mn@doi
  [\apj] {10.3847/1538-4357/abf8b8}, \href
  {https://ui.adsabs.harvard.edu/abs/2021ApJ...914...55W} {914, 55}

\bibitem[\protect\citeauthoryear{{Yuan} \& {Narayan}}{{Yuan} \&
  {Narayan}}{2014}]{YuanNarayan2014ARA&A..52..529Y}
{Yuan} F.,  {Narayan} R.,  2014, \mn@doi [\araa]
  {10.1146/annurev-astro-082812-141003}, \href
  {https://ui.adsabs.harvard.edu/abs/2014ARA&A..52..529Y} {52, 529}

\bibitem[\protect\citeauthoryear{Yuan, Gan, Narayan, Sadowski, Bu  \& Bai}{Yuan
  et~al.}{2015}]{yuan2015numerical}
Yuan F.,  Gan Z.,  Narayan R.,  Sadowski A.,  Bu D.,   Bai X.-N.,  2015, The
  Astrophysical Journal, 804, 101

\bibitem[\protect\citeauthoryear{Zhang, Shi, Shu  \& Zhou}{Zhang
  et~al.}{2003}]{Zhang2003PhysRevE.68.046709}
Zhang Y.-T.,  Shi J.,  Shu C.-W.,   Zhou Y.,  2003, \mn@doi [Phys. Rev. E]
  {10.1103/PhysRevE.68.046709}, 68, 046709

\bibitem[\protect\citeauthoryear{{Zhdankin}, {Ripperda}  \&
  {Philippov}}{{Zhdankin} et~al.}{2023}]{Zhdankin2023PhRvR...5d3023Z}
{Zhdankin} V.,  {Ripperda} B.,   {Philippov} A.~A.,  2023, \mn@doi [Physical
  Review Research] {10.1103/PhysRevResearch.5.043023}, \href
  {https://ui.adsabs.harvard.edu/abs/2023PhRvR...5d3023Z} {5, 043023}

\makeatother
\end{thebibliography}
